\documentclass[aps,pre,superscriptaddress,twocolumn,tightenlines,showpacs,floatfix,amsmath,amssymb,pdftex]{revtex4-1}

\usepackage{amsmath}
\usepackage{amssymb}
\usepackage{bm}
\usepackage{bbm}
\usepackage{float}
\usepackage{tikz} % for colorted text box
\usepackage{graphicx}
\usepackage{multirow}
\usepackage{booktabs}
\def\id{\mathbbm{1}}
\def\integers{\mathbb{Z}}
\newcommand{\corr}[1]{\langle #1\rangle}
\newcommand{\vek}[1]{\mathbf{#1}}

\newcommand{\f}[2]{\frac{#1}{#2}}
\newcommand{\de}{\mathrm{d}}

\newenvironment{matriisi}{\left(\begin{matrix}}{\end{matrix}\right)}

\begin{document}
\title{Classification of point-group-symmetric orientational ordering tensors}
\date{\today}

\author{Jaakko Nissinen}
\affiliation{Instituut-Lorentz for Theoretical Physics, Universiteit Leiden,
PO Box 9506, NL-2300 RA Leiden, The Netherlands}
\author{Ke Liu}
\email{ke@lorentz.leideinuniv.nl}
\affiliation{Instituut-Lorentz for Theoretical Physics, Universiteit Leiden,
PO Box 9506, NL-2300 RA Leiden, The Netherlands}
\author{Robert-Jan Slager}
\affiliation{Instituut-Lorentz for Theoretical Physics, Universiteit Leiden,
PO Box 9506, NL-2300 RA Leiden, The Netherlands}
\author{Kai Wu}
\email{kaiwu@stanford.edu}
\affiliation{Stanford Institute for Materials and Energy Sciences, SLAC National Accelerator Laboratory and Stanford University,
Menlo Park, California 94025, USA}
\author{Jan Zaanen}
\affiliation{Instituut-Lorentz for Theoretical Physics, Universiteit Leiden,
PO Box 9506, NL-2300 RA Leiden, The Netherlands}

\begin{abstract}
The concept of symmetry breaking has been a propelling force in understanding phases of matter. 
%While spatial symmetry breaking is one of the most prevalent examples, its rotational symmetries have nevertheless not been explored to full extent: apart from extensive studies of generic examples of rotational order, such as the uniaxial nematic phase associated with point group $D_{\infty h}$, the richness of the point group hierarchy descending from the full rotational group $O(3)$ has remarkably not been fully explored. 
While rotational symmetry breaking is one of the most prevalent examples, the rich landscape of orientational orders breaking the rotational symmetries of isotropic space, i.e. $O(3)$, to a three-dimensional point group remain largely unexplored, apart from simple examples such as ferromagnetic or uniaxial nematic ordering. 
Here we provide an explicit construction, utilizing a recently introduced gauge theoretical framework, to address the three-dimensional point-group-symmetric orientational orders on a general footing. This unified approach allows us to enlist order parameter tensors for all three dimensional point groups. By construction, these tensor order parameters are the minimal set of simplest tensors allowed by the symmetries that uniquely characterize the orientational order. We explicitly give these for the point groups $\{C_n, D_n, T, O, I\} \subset SO(3)$ and $\{C_{nv}, S_{2n}, C_{nh}, D_{nh}, D_{nd}, T_h, T_d, O_h, I_h\}\subset O(3)$ for $n, 2n \in \{1,2,3,4,6, \infty\}$. This central result may be perceived as a roadmap for identifying exotic orientational orders that may become more and more in reach in view of rapid experimental progress in e.g. nano-colloidal systems and novel magnets.
\end{abstract}

\maketitle

\section{Introduction}

\begin{figure}
\includegraphics[width=0.45\textwidth]{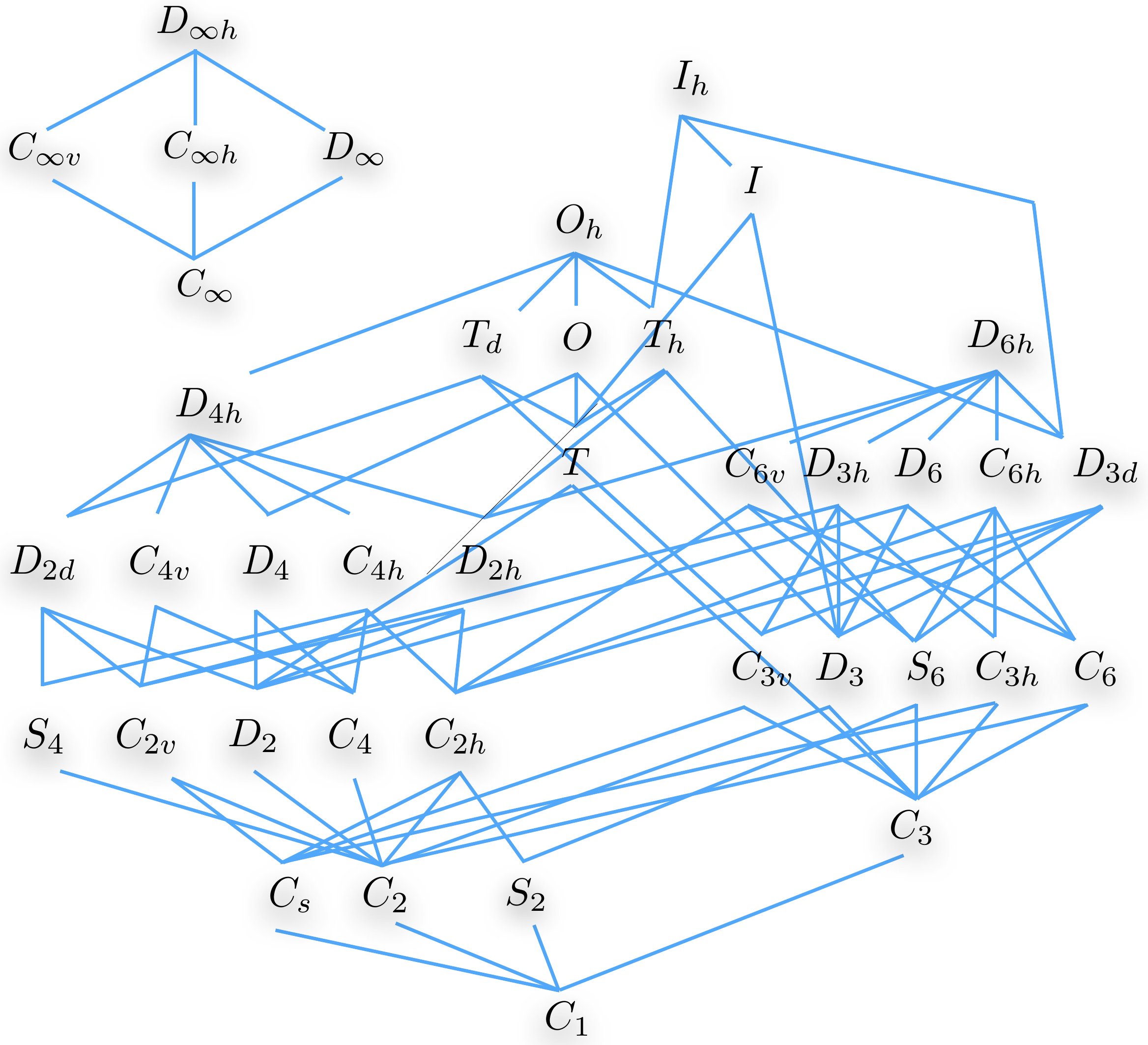}
\caption{A selection of finite subgroups of $O(3)$ or 3D point groups and their subgroup relations (not unique). They comprise continuous subgroups with a $C_{\infty} \simeq SO(2)$ axis and finite point groups. Ascending the hierarchy of symmetries, the order parameter tensor  describing the phases associated with the point group symmetries becomes increasingly complicated.}
\label{fig:subgroups}
\end{figure}

Rotational symmetry breaking of the three dimensional (3D) orthogonal group $O(3)$ plays an important role in many condensed matters systems, from classical and quantum spins to molecular and strongly correlated electronic nematic liquids  \cite{Anderson1984, DeGennesProst95, ChaikinLubensky00, FradkinKivelson99, FradkinEtAl10}. In familiar instances, like the Heisenberg ferromagnet and the uniaxial nematic, the full rotational group $O(3)$ is broken to $O(2)$ and $D_{\infty h}$, respectively.
However these are in fact only two special cases of the rich landscape of $O(3)$ symmetry breaking.
%a much richer landscape containing a vast number of possibilities.
Indeed, as a matter of principle, matter can break the rotational symmetries of isotropic space $O(3)$ to any of its subgroups, leading to long range orientational order characterized by complicated tensors order parameters. The subgroups of $O(3)$ have been mathematically identified for a long time and entail the well-known three-dimensional point groups. In Fig. \ref{fig:subgroups}, we show a representative selection of these symmetry groups and moreover their characterizing symmetry hierarchy.

Perhaps surprisingly, it appears that the zoo of  point-group orientational orders has never been explored in full generality.
Needless to say, the classification of rotational order parameters for some non-broken symmetries has been gradually accumulating since the past century due to various motives. 
% Needless to say, the classification of rotational order parameters for the two simplest and related examples has been gradually accumulating since the past century due to various motives. 
Firstly, in the soft matter literature the unixial ($D_{\infty h}$)  and biaxial ($D_{2h}$) order parameter have been shown to be characterized by second-rank tensors \cite{DeGennesProst95}, which have been intensively studied in various theories \cite{ Freiser70, Alben73, Straley74, BiaxialBook2015, AllenderLeeHafiz85, BiscariniZannoni1995, Rosso2007, AllenderLonga08, Greco2014}.
In addition, higher rank ordering tensors for the $T_d$-tetrahedral \cite{Fel1995, RadzihovskyLubensky2001, LubenskyRadzihovsky2002, Romano08}, $O_h$-cubic \cite{NelsonToner81, Jaric86, Romano06} and $I_h$-icosahedral \cite{SteinhardtNelson1981, SteinhardtNelson1983, Jaric1985, Jaric86, Fel1995, HajiAkbariGlotzer15} orders have been discussed by many authors e.g. in the context of Landau theories and nematic lattice models. 
Nonetheless, these cover still only a small subset of all 3D point group symmetries and, to the best of our knowledge, the order parameters for most instances are not known explicitly nor have appeared within a single unified classification scheme. These general order parameters, however, are becoming of more practical interest. 
New exotic orientational orders may be realized in ensembles of anisotropic constituents, especially nano- and colloidal particles of different shapes \cite{DamascenoGlotzer2012, vanAndersGlotzer2014}. 
In particular, the increasing experimental ability to control such degrees of freedom \cite{Glotzer2007, SunXia2002, HuangMichael2012, Mark2013} is especially promising in this regard.
Secondly, many unconventional orientational orders have also been proposed for quantum magnets \cite{MessioLhuillierMisguich11, GroverSenthil2011, XuLudwig2012} and spinor condensates  \cite{BarnettDemler2006, ZiboldEtAl16}.
In all these cases, the order parameters associated with each symmetry are indispensable to eventually verify the symmetry of these phases and the associated physics.

In this paper we bridge this divide and set out to classify the order parameters associated with arbitrary point groups within a single systematic framework. In particular, we adopt a gauge theoretical formulation, which we recently proposed in the study of generalized nematics \cite{LiuEtAl2015b}. Here we derive the full table of order parameters for physically most relevant 3D point groups. In detail, we highlight the order parameters for physically interesting symmetries including all the crystallographic point groups, the icosahedral groups arising in the context of quasi-crystalline ordering, and the five infinite axial groups $\{C_{\infty} \simeq SO(2), C_{\infty v } \simeq O(2), C_{\infty h}, D_{\infty}, D_{\infty h}\}$ exhibiting a continuous rotational $SO(2)$-axis.
We show that in order to uniquely characterize a point-group-symmetric orientational order of a phase, at most two order parameter tensors and a pseudoscalar are needed:
the second ordering tensor is required by the finite axial groups $\{C_n, C_{nv}, S_{2n}, C_{nh}, D_{n}, D_{nh}, D_{nd} \}$, whereas the pseudoscalar chiral order parameter is a requisite associated with the handedness or chirality of the proper point groups $\{C_n, D_n, T, O, I \}$ that are subgroups of the group of proper three-dimensional rotations $SO(3)$.

This paper is organized as follows. In Section \ref{sec:ops}, we discuss the general framework and present our main result the classification of  the order parameter tensors for three-dimensional point groups and discuss their relation to the subgroup structure.
In Section \ref{sec:gaugetheory}, we review the gauge theory description of generalized nematics and show it can act as an efficient order parameter generator.
The procedure of deriving the order parameters is demonstrated in Section \ref{sec:deriving_ops} and we discuss how to measure these order parameters in simulations from a more practical view point in Section \ref{sec:measurement}. We discuss many examples in Section \ref{sec:examples} before concluding in Section \ref{sec:summary} with an outlook.

\section{Orientational ordering tensors} \label{sec:ops}

In this Section we announce our main result in Table \ref{table:ops}. In order for the reader to able to read Table \ref{table:ops}, we first discuss the general problem of determining order parameter tensor, then specify our conventions and discuss the broad classifications of in terms of the structure of three-dimensional point groups.

\subsection{Three-dimensional orientational ordering tensors}

Let us begin by recalling the characterization of rotational ordering in the familiar context of the Heisenberg ferromagnet and the conventional uniaxial nematic. 
In the ferromagnetic phase of a classical Heisenberg magnet, the rotational $O(3)$ symmetry of the Hamiltonian breaks down to the point group $C_{\infty v} \simeq O(2)$ defined by the axis of magnetization $\vek{M}$. The order parameter $\vek{M} = \langle \mathbf{n}_i \rangle$ is given by the macroscopic averaging of local spins $\mathbf{n}_i$ and is a 3D vector with an orientational order parameter space $O(3)/O(2)\simeq S^2$.
On the other hand, for uniaxial liquid crystals or spin nematics, where the $O(3)$ symmetry is broken to the point group $D_{\infty h}$ in the ordered phase, the system exhibits a macroscopic ordering along an axis $\vek{n}$. The uniaxial symmetry $D_{\infty h}$ acts on the order parameter as $\vek{n}\to -\vek{n}$ and these describe the same macroscopic ordering.  Often depicted as being formed of explicitly rod-like ``molecules", a coarse-grained order parameter can be formulated in terms of a local vector $\vek{n}_i$ along the ``long" axis of each ``molecule", with the identification of $\mathbf{n}_i$ with $-\mathbf{n}_i$. To define the uniaxial orientational order, one therefore needs a second rank tensor, $\mathbb{Q}[\vek{n}] =  \mathbf{n} \otimes \mathbf{n} - \frac{\id}{3}$ %= \corr{\vek{n}_i\otimes \vek{n}_i-\frac{\id}{3}}$,
 which is characterized by its invariance under $\mathbf{n} \rightarrow -\mathbf{n}$ and is zero in the isotropic phase by being traceless. Accordingly, the global order parameter is defined as $\mathbb{Q}[\vek{n}] = \langle \mathbb{Q}[\vek{n}_i] \rangle$ in the coarse-grained order parameter theory and formally relates to the unixial order parameter space $O(3)/D_{\infty h} \simeq S^2/\integers_2 \simeq \mathbb{RP}^2$, the real projective plane. 

The above familiar examples share the key feature of having an $O(2)$ symmetry in the plane perpendicular to the ordering vector, which is why the underlying physics is so apparent: the order parameter is defined by one axis and the rotations in the perpendicular plane are trivial, and the degrees of freedom effectively reduce to 1D objects (the spins and the rods in the above examples). Nonetheless, for general 3D point-group-symmetric ordering, the order parameter and the coarse grained degrees of freedom form intrinsic 3D objects. To define the 3D orientation one therefore has to depart from a full $O(3)$ rotation matrix $R$,
\begin{align}
R = \big( \mathbf{l} \quad \mathbf{m} \quad \mathbf{n} \big)^T.
\end{align} 
The rows $ \{ \mathbf{l}, \mathbf{m}, \mathbf{n} \}$ of $R$ form an orthonormal triad  $\vek{n}^{\alpha}= \{\vek{l},\vek{m},\vek{n}\}$. 
In other words, $R$ is a rotation that brings the triad $\vek{n}^{\alpha} =\{\vek{l},\vek{m},\vek{n} \}$ into coincidence with a fixed ``laboratory" frame $\mathbf{e}_a = \{ \mathbf{e}_1, \mathbf{e}_2, \mathbf{e}_3\}$ and can be defined by three Euler angles with respect to the unit vectors $\mathbf{e}_a$.
The determinant of $R$ defines the handedness or chirality of the triad,
\begin{equation} \label{eq:sigma}
\sigma = \det R = \epsilon_{abc}(\vek{l}\otimes \vek{m} \otimes \vek{n})_{abc} =  \mathbf{l} \cdot ( \mathbf{m} \times \mathbf{n} ) = \pm 1,
\end{equation}
which is a pseudoscalar and invariant under the proper rotations $SO(3)$. Moreover, due to $O(3) = SO(3) \times \{\id, -\id \}$, we have the decomposition
\begin{align}
R = \sigma\widetilde{R} = \sigma(\widetilde{\vek{l}} \quad \widetilde{\vek{m}} \quad \widetilde{\vek{n}})^T
\end{align}
where $\widetilde{R} \in SO(3)$ and its rows $\tilde{\vek{n}}^{\alpha} = \{\tilde{\vek{l}},\tilde{\vek{m}},\tilde{\vek{n}}\}$ are pseudovectors. The $O(3)$ constraints $R^T R = RR^T = \id$ and $\det R = \pm 1$ of course reduce the free parameters to the three Euler angles $\Omega = (\theta, \phi, \varphi)$ and chirality in the frame $\vek{e}_a$ but we will find the vector notation with the $O(3)$-constraints understood very useful in the following.

%As a result, the general approach is clear by generalization and an orientational order parameter with a point group symmetry $G$ is now defined by a $G$-invariant tensor. One immediately realizes, however, that the rotation matrix $R$ or the triad $\vek{n}^{\alpha} = \{\vek{l},\vek{m},\vek{n}\}$ cannot serve this purpose but higher order tensors are needed in general and can be constructed from $R$ in terms of the triad $\vek{n}^{\alpha}$. 
In the general case, the order parameter has to be invariant under all unbroken point-group transformations. As a result, an orientational order parameter with a point group symmetry $G$ is defined by $G$-invariant tensors constructed from the triad $R$ or $\vek{n}^{\alpha}= \{\vek{l},\vek{m},\vek{n}\}$.
These tensors are equivalent to higher order multipoles or (three-dimensional) spherical harmonics. We will denote these order parameters tensors composed of the triads generically as $\mathbb{O}^{G}$, where the additional label specifies the symmetry group $G$ when appropriate.
Concretely, in the above two examples, the order parameter tensor is the magnetization vector $\mathbb{O}^{C_{\infty v}}[\vek{n}] = \mathbf{n}$ and the second rank tensor or director $\mathbb{O}^{D_{\infty h}}[\vek{n}] = \mathbb{Q}[\vek{n}]$, respectively.
Finally, we note that besides the orientational order, the \emph{composite} chiral order parameter $\sigma$ defined in Eq. \eqref{eq:sigma} is needed for proper point-group symmetry such as $\{C_n, D_n, T, O, I \}$ due to the breaking of the chiral symmetry of $O(3)$. We also note that $O(3)$-triad order parameters constructed from $R$ effectively also arise e.g. in spin-models with non-homogenous ordering associated with sublattice structures \cite{MessioLhuillierMisguich11}.

As $\mathbb{O}^{G}$ needs to be uniquely invariant under a given symmetry $G$ in the hierarchy in Fig. \ref{fig:subgroups}, it is in general highly non-trivial to construct its explicit form, even though the polynomial invariants of 3D points groups have been computed a long time ago \cite{JaricSharp1984, Michel2001} and the representation theory of $SO(3)$ is known. As a central result, we will discuss a systematic and straightforward framework to derive these order parameter tensors in Sections \ref{sec:gaugetheory} and \ref{sec:deriving_ops}, the result of which is shown in Table \ref{table:ops}. 
 We show the lowest order invariant tensors $\mathbb{O}^G$ for several point groups, where $\mathbb{O}^G = \mathbb{O}^G[\vek{l}, \vek{m}, \vek{n}]$ for polyhedral groups; $\mathbb{O}^G = \{ \mathbb{A}^G, \mathbb{B}^G \}$ for axial groups,
where $\mathbb{A}^G = \mathbb{A}^G[\vek{n}]$ is  the order parameter for the main axis $\mathbf{n}$ and $\mathbb{B}^G = \mathbb{B}^G[\mathbf{l}, \mathbf{m}]$ or  $\mathbb{B}^G[\mathbf{l}, \mathbf{m}, \mathbf{n}]$ denote secondary order parameters defining the in-plane structure for the finite axial groups.
 We note that amongst the ordering tensors in Table \ref{table:ops}, the $C_1$ order parameters $\mathbb{O}^{C_{1}}[ \vek{l},\vek{m},\vek{n} ] = \{ \mathbb{A}^{C_{\infty v}}[\vek{n}], \mathbb{B}^{C_1}[\vek{l},\vek{m}] \} = \{\vek{l},\vek{m},\vek{n} \}$ simply constitute the original $O(3)$-rotor order parameter $R$ of a phase with no unbroken symmetry ($C_1$ is the trivial group);
 $\mathbb{O}^{D_{2h}} = \{ \mathbb{O}^{D_{\infty h}}[\vek{n}], \mathbb{B}^{D_{2h}}[\vek{l},\vek{m}] \}$ compose the well known order parameter tensors for $D_{2h}$-biaxial nematics; 
 $\mathbb{O}^{C_{\infty v}}[\vek{n}]$ and $\mathbb{O}^{D_{\infty h}}[\vek{n}]$ are the classical Heisenberg spin $\vek{n}$ and uniaxial director $\mathbb{Q}[\vek{n}]$, respectively; $\mathbb{O}^{O_h}[\vek{l},\vek{m},\vek{n}]$ has been discussed in Ref. \cite{NelsonToner81}; $\mathbb{O}^{T_d}[\vek{l},\vek{m},\vek{n}]$ and  $\mathbb{O}^{I_h}[\vek{l},\vek{m},\vek{n}]$ appear in a different form in Ref. \cite{Fel1995}, where an incomplete classification of order parameters for subgroups of $SO(3)$ is also discussed. In a recent paper \cite{HajiAkbariGlotzer15}, a classification of unique set of order parameters for subgroups of $SO(3)$ was carried out. These order parameters, however, are not the most elementary (``strongest" in the terminology of Ref. \onlinecite{HajiAkbariGlotzer15}) set of order parameters possible, in the sense that for individual order parameters, when allowed by symmetry, $\mathbb{O}^{G} \neq 0$ does not necessary imply $G$ order but also $G'$-ordering for $G\subset G'$ or have the lowest possible rank. On the other hand, the gauge theoretical approach naturally provides such minimal order parameters allowed by the symmetries, as we highlight in Section \ref{sec:gaugetheory}. In summary, many of the order parameter tensors in Table \ref{table:ops} are new and have not been classified in the context of a single unified framework for all rotational symmetries of $O(3)$.

\setlength{\tabcolsep}{5pt}
\renewcommand{\arraystretch}{1.1}
\begin{table*}
\centering
\caption{\textbf{Invariant ordering tensors for three-dimensional point groups.} The first column specifies the symmetries and the second column specifies the type $\{\mathbb{O}, \mathbb{A}, \mathbb{B} \}$ of the ordering tensor. The third column gives the explicit form of the tensors. They are traceless and vanish in the isotropic phase but can have linear dependence, see Sec. \ref{sec:examples}. The infinite axial groups $\{C_{\infty}, C_{\infty v}, C_{\infty h}, D_{\infty}, D_{\infty h} \}$ require a single ordering tensor, $\mathbb{A}[\mathbf{n}]$, describing the orientation of their primary symmetry axis, chosen to be $\mathbf{n}$; the finite axial groups $\{ C_n, C_{nv}, C_{nh}, S_{2n}, D_n, D_{nh}, D_{nd}\}$ require two ordering tensors, $\mathbb{A}[\mathbf{n}]$ and $\mathbb{B}[\mathbf{l}, \mathbf{m}]$ or $ \mathbb{B}[\mathbf{l}, \mathbf{m}, \mathbf{n}]$, for their primary axis and perpendicular in-plane structure, respectively; the polyhedral groups $\{T, T_d, T_h, O, O_h, I, I_h\}$, which treat $\{ \mathbf{l}, \mathbf{m}, \mathbf{n}\}$ symmetrically, require only one ordering tensor $\mathbb{O}[\mathbf{l}, \mathbf{m}, \mathbf{n}]$. Due to the symmetry hierarchy, many point groups share ordering tensors (see Section \ref{sec:op_group}).
The handedness or chirality $\sigma = \det R =\pm 1$ arises for proper point groups. These ordering tensors uniquely define the orientational ordering associated with the point group symmetries.
For example, the order parameters for finite proper axial groups are given by $\mathbb{O}^G = \{ \mathbb{A}^G, \mathbb{B}^G, \sigma\}$. $^{\otimes n}$ denotes the tensor power, e.g., $ \mathbf{n}^{\otimes 2}=\vek{n}\otimes\vek{n}$ and $\delta_{ab}  \bigotimes_{ \mu =  a,b} \mathbf{e}_{\mu} = \delta_{ab} \mathbf{e}_a \otimes \mathbf{e}_b$. $\tau=(1+\sqrt{5})/2$ is the golden ratio.  $\sum_{\text{cyc}}$ runs over cyclic permutations of $\{ \mathbf{l}, \mathbf{m}, \mathbf{n} \}$. $\sum_{\text{pairs}}$ sums over all non-equivalent pairings of the indices of the Kronecker deltas. $\sum_{\text{pairs}^{\prime\prime}}$ in the $\{ C_{6v}, D_6, D_{6h} \}$ cases sums over all the six permutations of the indices $d$, $e$ and $f$, and $\sum_{\text{pairs}^{\prime}} = \sum_{\text{pairs}} - \sum_{\text{pairs}^{\prime\prime}}$. $\sum_{\{+,-\}}$ for the $\{ I, I_h \}$ is a sum over the four combinations of the two $\pm$ signs.} \label{table:ops}
\begin{tabular}{ | c | c | c |  c |  }
    \hline
    \hline
   \parbox{1.5cm}{\centering{\bf Symmetry \\ Groups}} 
	& \parbox{1.5cm}{\centering{\bf Type }}   
   & {\bf Ordering Tensors}  & \parbox{1.2cm}{\centering{\bf Tensor \\ Rank}}  \\ \hline
   
   $C_1$, $C_{1h}$ 
   & $\mathbb{B}[\mathbf{l}, \mathbf{m}]$
   & $\mathbf{l}$, $\mathbf{m}$ 
   & 1	
   \\ \hline
   
   $S_{2}$ 
    & $\mathbb{B}[\mathbf{l}, \mathbf{m}, \mathbf{n}]$
   & $ \mathbf{l} \otimes \mathbf{m}  $, $ \mathbf{m} \otimes \mathbf{l}  $, 
   		$ \mathbf{m} \otimes \mathbf{n} $, $ \mathbf{n} \otimes \mathbf{m} $,
   		$\mathbf{n} \otimes \mathbf{l}$, $\mathbf{l} \otimes \mathbf{n}$  
   & 2	
   \\ \hline
   
	$C_2$, $C_{2h}$ 
	 & $\mathbb{B}[\mathbf{l}, \mathbf{m}]$
	& $ \mathbf{l} \otimes \mathbf{m}  $, $ \mathbf{m} \otimes \mathbf{l}  $
	& 2	
	\\ \hline   
   
   $C_{2v}$, $D_2$, $D_{2h}$ 
    & $\mathbb{B}[\mathbf{l}, \mathbf{m}]$
   & $\mathbf{l} \otimes \mathbf{l} - \frac{1}{3} \id$,  $ \mathbf{m} \otimes \mathbf{m} - \frac{1}{3} \id$
   & 2
   \\ \hline
   
   $S_4$
    & $\mathbb{B}[\mathbf{l}, \mathbf{m}, \mathbf{n}]$
   & $  \big(\mathbf{l} \otimes \mathbf{l} - \mathbf{m}\otimes \mathbf{m}\big) \otimes \mathbf{n}$
   & 3
   \\ \hline
   
   $D_{2d}$
    & $\mathbb{B}[\mathbf{l}, \mathbf{m}, \mathbf{n}]$
   & $  \big(\mathbf{l} \otimes \mathbf{m} + \mathbf{m} \otimes \mathbf{l} \big) \otimes \mathbf{n}$
   & 3
   \\ \hline
   
   $C_3$, $C_{3h}$
    & $\mathbb{B}[\mathbf{l}, \mathbf{m}]$
   & \parbox{10cm}{\centering
   		$ \big( \mathbf{l}^{\otimes 3} - \mathbf{l} \otimes \mathbf{m}^{ \otimes 2} - \mathbf{m} \otimes \mathbf{l} \otimes \mathbf{m} - \mathbf{m}^{ \otimes 2} \otimes \mathbf{l}\big)$,
    	\\
      $ \big( \mathbf{m}^{\otimes 3} - \mathbf{m} \otimes \mathbf{l}^{ \otimes 2} - \mathbf{l} \otimes \mathbf{m} \otimes \mathbf{l} - \mathbf{l}^{ \otimes 2} \otimes \mathbf{m}\big)$ }
   & 3
   \\ \hline

   $C_{3v}$, $D_3$, $D_{3h}$
    & $\mathbb{B}[\mathbf{l}, \mathbf{m}]$
   & $ \big( \mathbf{l}^{\otimes 3} - \mathbf{l} \otimes \mathbf{m}^{ \otimes 2} - \mathbf{m} \otimes \mathbf{l} \otimes \mathbf{m} - \mathbf{m}^{ \otimes 2} \otimes \mathbf{l}\big)$
   & 3
   \\ \hline 
   
   $S_6$
    & $\mathbb{B}[\mathbf{l}, \mathbf{m}, \mathbf{n}]$
   & \parbox{10cm}{\centering
   		$ \big( \mathbf{l}^{\otimes 3} - \mathbf{l} \otimes \mathbf{m}^{ \otimes 2} - \mathbf{m} \otimes \mathbf{l} \otimes \mathbf{m} - \mathbf{m}^{ \otimes 2} \otimes \mathbf{l}\big) \otimes \mathbf{n}$,
    	\\ 
      $ \big( \mathbf{m}^{\otimes 3} - \mathbf{m} \otimes \mathbf{l}^{ \otimes 2} - \mathbf{l} \otimes \mathbf{m} \otimes \mathbf{l} - \mathbf{l}^{ \otimes 2} \otimes \mathbf{m} \big) \otimes \mathbf{n}$ }
   & 4
   \\ \hline

   $D_{3d}$
   & $\mathbb{B}[\mathbf{l}, \mathbf{m}, \mathbf{n}]$
   & $ \big( \mathbf{m}^{\otimes 3} - \mathbf{m} \otimes \mathbf{l}^{ \otimes 2} - \mathbf{l} \otimes \mathbf{m} \otimes \mathbf{l} - \mathbf{l}^ {\otimes 2} \otimes \mathbf{m} \big) \otimes \mathbf{n}$ 
   & 4 
   \\ \hline
   
   $C_4$, $C_{4h}$
   & $\mathbb{B}[\mathbf{l}, \mathbf{m}]$
   & $\mathbf{l}^{\otimes 3} \otimes \mathbf{m} - \mathbf{m}^{\otimes 3} \otimes \mathbf{l}$
   & 4
   \\ \hline
   
   $C_{4v}$, $D_4$, $D_{4h}$
   & $\mathbb{B}[\mathbf{l}, \mathbf{m}]$
   & \parbox{10cm}{\centering
   $\mathbf{l}^{\otimes 2} \otimes \mathbf{m} ^{\otimes 2} + \mathbf{m}^{\otimes 2} \otimes \mathbf{l} ^{\otimes 2} 
   -\frac{4}{15}  \delta_{ab} \delta_{cd} \bigotimes_{\substack{ \mu =  a,b,c, d}} \mathbf{e}_{\mu}
   + \frac{1}{15} \big(  \delta_{ac} \delta_{bd} \bigotimes_{\substack{ \mu =  a,c,b, d}} \mathbf{e}_{\mu}
    +  \delta_{ad} \delta_{bc}\bigotimes_{\substack{ \mu =  a,d,b, c}} \mathbf{e}_{\mu}  \big)$,
$  \mathbf{l}^{\otimes 4}  +  \mathbf{m}^{\otimes 4}  
-\frac{2}{15}\sum_{\text{pairs}} \delta_{ab} \delta_{cd} \bigotimes_{\substack{ \mu =  a,b,c, d}} \mathbf{e}_{\mu}$        
    }
   & 4 
   \\ \hline
   
   $C_6$, $C_{6h}$
   & $\mathbb{B}[\mathbf{l}, \mathbf{m}]$
   &  \parbox{10cm}{\centering
   		$ \big( \mathbf{l}^{\otimes 3} - \mathbf{l} \otimes \mathbf{m}^{ \otimes 2} - \mathbf{m} \otimes \mathbf{l} \otimes \mathbf{m} - 	\mathbf{m}^{ \otimes 2} \otimes \mathbf{l}\big)
   		\otimes \big( \mathbf{m}^{\otimes 3} - \mathbf{m} \otimes \mathbf{l}^{ \otimes 2} - \mathbf{l} \otimes \mathbf{m} \otimes \mathbf{l} - \mathbf{l}^{ \otimes 2} \otimes \mathbf{m}\big)$, $\big( \mathbf{m}^{\otimes 3} - \mathbf{m} \otimes \mathbf{l}^{ \otimes 2} - \mathbf{l} \otimes \mathbf{m} \otimes \mathbf{l} - \mathbf{l}^{ \otimes 2} \otimes \mathbf{m}\big)
       \otimes  \big( \mathbf{l}^{\otimes 3} - \mathbf{l} \otimes \mathbf{m}^{ \otimes 2} - \mathbf{m} \otimes \mathbf{l} \otimes \mathbf{m} - 	\mathbf{m}^{ \otimes 2} \otimes \mathbf{l}\big)$ }   
   
   & 6
   \\ \hline
   
   $C_{6v}$, $D_{6}$, $D_{6h}$
   & $\mathbb{B}[\mathbf{l}, \mathbf{m}]$
   &  \parbox{10cm}{\centering
   		$  \big( \mathbf{l}^{\otimes 3} - \mathbf{l} \otimes \mathbf{m}^{ \otimes 2} - \mathbf{m} \otimes \mathbf{l} \otimes \mathbf{m} - 	\mathbf{m}^{ \otimes 2} \otimes \mathbf{l}\big)^{\otimes 2}
   		 + \frac{4}{105} \sum_{\text{pairs}^{\prime}} \delta_{ab} \delta_{cd} \delta_{ef}  \bigotimes_{\substack{ \mu =  a,b,c, \\  d,e,f}} \mathbf{e}_{\mu} 
      -\frac{2}{21} \sum_{\text{pairs}^{\prime \prime}} 
\delta_{ad} \delta_{be} \delta_{cf}       
      \bigotimes_{\substack{ \mu =  a,d,b, \\  e,c,f}} \mathbf{e}_{\mu}  $,
$\big( \mathbf{m}^{\otimes 3} - \mathbf{m} \otimes \mathbf{l}^{ \otimes 2} - \mathbf{l} \otimes \mathbf{m} \otimes \mathbf{l} - \mathbf{l}^{ \otimes 2} \otimes \mathbf{m}\big)^{\otimes 2}
+ \frac{4}{105} \sum_{\text{pairs}^{\prime}} \delta_{ab} \delta_{cd} \delta_{ef}  \bigotimes_{\substack{ \mu =  a,b,c, \\  d,e,f}} \mathbf{e}_{\mu} 
      -\frac{2}{21} \sum_{\text{pairs}^{\prime \prime}} 
\delta_{ad} \delta_{be} \delta_{cf}       
      \bigotimes_{\substack{ \mu =  a,d,b, \\  e,c,f}} \mathbf{e}_{\mu}  $  }     
   & 6
   \\ \hline
   
   \parbox{3cm}{\centering $C_n, C_{nv}, C_{\infty} , C_{\infty v}$ }
   & $\mathbb{A}[\mathbf{n}]$
   & $\mathbf{n}$
   & 1
   \\ \hline
   
   $C_{\infty h}$
      & $\mathbb{A}[\mathbf{n}]$
   & $\sigma \mathbf{n}$
   & 1
   \\ \hline
      
   \parbox{3.cm}{\centering $S_{2n}, C_{nh}, D_n, $ \\ $D_{nh}, D_{nd}, D_{\infty}$, $D_{\infty h}$}
      & $\mathbb{A}[\mathbf{n}]$
   & $ \mathbf{n} \otimes \mathbf{n} - \frac{1}{3} \id$ 
   & 2
   \\ \hline 
   
   $T$
      & $\mathbb{O}[\mathbf{l}, \mathbf{m}, \mathbf{n}]$
   & $ \sum_{\text{cyc}} \mathbf{l} \otimes \mathbf{m} \otimes \mathbf{n}$
   & 3
   \\ \hline
   
   $T_d$
   & $\mathbb{O}[\mathbf{l}, \mathbf{m}, \mathbf{n}]$
   & $   \sum_{\text{cyc}} \big ( \mathbf{l} \otimes \mathbf{m}  + \mathbf{m} \otimes \mathbf{l} \big) \otimes \mathbf{n}$
   & 3
   \\ \hline
     
   $T_h$
   & $\mathbb{O}[\mathbf{l}, \mathbf{m}, \mathbf{n}]$
   & \parbox{10cm}{\centering
   $\mathbf{m}^{\otimes 2} \otimes \mathbf{l}^{\otimes 2} +  \mathbf{l}^{\otimes 2} \otimes \mathbf{n}^{\otimes 2} +  \mathbf{n}^{\otimes 2} \otimes \mathbf{m}^{\otimes 2}-\frac{2}{5} \delta_{ab} \delta_{cd}\bigotimes_{\substack{ \mu =  a,b,c, d}} \mathbf{e}_{\mu}
+ \frac{1}{10} \big( \delta_{ac} \delta_{bd} \bigotimes_{\substack{ \mu =  a,c,b, d}} \mathbf{e}_{\mu}
    + \delta_{ad} \delta_{bc}\bigotimes_{\substack{ \mu =  a,d,b, c}} \mathbf{e}_{\mu} \big)$ }
   & 4
   \\ \hline
   
   $O$, $O_h$
   & $\mathbb{O}[\mathbf{l}, \mathbf{m}, \mathbf{n}]$
   &  $  \mathbf{l}^{\otimes 4}  +  \mathbf{m}^{\otimes 4} + \mathbf{n}^{\otimes 4} 
-\frac{1}{5}\sum_{\text{pairs}} \delta_{ab} \delta_{cd} \bigotimes_{\substack{ \mu =  a,b,c, d}} \mathbf{e}_{\mu}$
   & 4 
   \\ \hline
   
   $I$, $I_{h}$
   & $\mathbb{O}[\mathbf{l}, \mathbf{m}, \mathbf{n}]$
   & $ \sum_{\text{cyc}} \big [ \mathbf{l}^{\otimes 6} +\sum_{ \{+,- \}} 
			\big(\frac{1}{2} \mathbf{l} \pm \frac{\tau}{2} \mathbf{m} \pm \frac{1}{2\tau} \mathbf{n} \big)^{\otimes 6} \big]
  		-\frac{1}{7} \sum_{\text{pairs}} \delta_{ab} \delta_{cd} \delta_{ef} \bigotimes_{\substack{ \mu =  a,b,c, \\  d,e,f}} \mathbf{e}_{\mu}  
 	  $
  & 6
    \\ \hline   
   \hline 
\end{tabular}

\end{table*}

\subsection{Point groups and invariant tensors}\label{sec:class}

In Table \ref{table:ops}, where the standard Sch\"onflies notation is used  \cite{SternbergBook, Michel2001}, we show the invariant tensors from which suitable order parameter tensors can be simply constructed for all the $32$ crystallographic point groups, the $2$ icosahedral groups and the $5$ infinite axial groups.
These tensors can be divided into three classes in terms of the structure of point group symmetries. Together with the handedness field $\sigma$, they can uniquely define the order parameter for the symmetries mentioned above. 

Throughout this paper, we will refer to rotations $R\in SO(3)$ as \emph{proper} and all elements $R\in O(3)$ with $\det R = -1$ as \emph{improper} and use the analogous nomenclature concerning the subgroups of $O(3)$. We first note the following well-known structure of the subgroups of $O(3) = SO(3)\times \{\id, -\id\}$:
\begin{itemize}
\item[(i)] Proper point groups $\{C_n, D_n, T, O,I\}$ that are subgroups of $SO(3)$.

\item[(ii)] Subgroups of the form $G=H\times \{\id, -\id \}$ for $H\subset SO(3)$, including axial groups $\{ C_{nh}, D_{nh}\}$ for even and infinite $n$ ($C_{\infty h} \simeq S_{\infty}$, $D_{\infty d} \simeq D_{\infty h} $), $\{S_{2n}, D_{nd} \}$ for odd $n$ and polyhedral groups $\{ T_h, O_h, I_h\}$.

\item[(iii)] Subgroups $G$ with improper rotations that do not contain $-\id$, including $C_{nv}$ for all $n$,  $\{ C_{nh}, D_{nh}\}$ for odd $n$, $\{S_{2n}, D_{nd} \}$  for even $n$ and the regular tetrahedral group $T_d$.
These groups are of the form $G= H \cup -(G' \setminus H)$, where $H$ is a normal subgroup of index two for $G' \subset SO(3)$ \cite{SternbergBook,Michel2001}.
\end{itemize}

Our point groups are defined in the coordinate system spanned by the unit triad vectors $\vek{n}^{\alpha} = \{\vek{l},\vek{m},\vek{n}\}$ set up in the following way. All point groups have the origin as their fixed point. 
The rotational axis of cyclic rotation groups $C_n$ of is chosen to be $\vek{n}$. The dihedral group $D_n$ has an additional generator in terms of a $\pi$-rotation along the vector $\vek{l}$ (or $\mathbf{m}$). The group $C_{nv}$ is augmented with a ``vertical" reflection in the plane $(\vek{l},\vek{n})$. The groups $C_{nh}$ and $D_{nh}$ have an additional ``horizontal" reflection plane $(\vek{l},\vek{m})$. The group $D_{nd}$ has vertical reflection planes in terms of bisectors of the dihedral $\pi$-rotation axes. The groups $S_{2n}$ are composed of $n$-fold rotoreflections in the plane $\vek{l},\vek{m}$. The polyhedral  groups $T,T_d, T_h$ and $O, O_h$ are defined in terms of a (tetrahedron embedded) in  a cube with face normals $\vek{n}^{\alpha}= \{\vek{l},\vek{m},\vek{n}\}$. The group $I_h$ is the symmetry group of an icosahedron with vertices at cyclic permutations of the coordinates $\pm \tau \vek{l} \pm \vek{m} \pm 0 \cdot \vek{n}$ and $I$ its proper subgroup, following the conventions in \cite{Litvin91}.

In Table \ref{table:ops}, we have collected the simplest invariant tensors for those point groups in the coordinate system discussed above. Let us now explain the broad characteristics of Table \ref{table:ops} focusing on the nature of the groups according to Fig. \ref{fig:subgroups}.

\subsubsection{Continuous axial groups}
The five infinite axial groups $\{ C_{\infty}, C_{\infty v}, C_{\infty h}, D_{\infty}, D_{\infty h}\}$ require only one tensor to define the associated orientational order. 
This is because these groups contain a plane perpendicular to the vector $\vek{n}$ with continuous $SO(2)$ or $O(2)$ rotations, hence their in-plane structure is trivial and the order parameter effectively reduces to a vector ($C_{\infty}, C_{\infty v}$), a pseudovector ($C_{\infty h}$) or a director ($D_{\infty}, D_{\infty h}$), up to an additional chiral order parameter $\sigma$ for the proper point groups.

\subsubsection{Finite axial groups}
Finite axial groups $\{ C_{n}, C_{nv}, S_{2n}, C_{nh}, D_n, D_{nh}, D_{nd}\}$ require two ordering tensors $\{\mathbb{A}, \mathbb{B} \}$: $\mathbb{A} = \mathbb{A}[\vek{n}]$ describes the orientation of the primary axis, which is always chosen as $\mathbf{n}$ in Table \ref{table:ops}, 
and tensors $\mathbb{B} = \mathbb{B}[\vek{l},\vek{m}]$ or $\mathbb{B}[\vek{l},\vek{m},\vek{n}]$ for the perpendicular in-plane order. This generalizes well-known structure of the order parameters of biaxial ($D_{2h}$) liquid crystals.
Due to symmetry relations which will be discussed later, the primary ordering tensors $\mathbb{A}[\vek{n}]$ for $\{ C_{n}, C_{nv},\}$ and $\{ S_{2n}, C_{nh}, D_n, D_{nh}, D_{nd}\}$ are identical to the order parameters $\mathbb{O}^{C_{\infty v}}[\vek{n}]$ and $\mathbb{O}^{D_{\infty h}}[\vek{n}]$, respectively.

\subsubsection{Polyhedral groups} 
The finite symmetry groups  $\{ T, T_d, T_h, O, O_h, I, I_h\}$ related to the regular tetrahedron, octahedron and icosahedron, respectively, require only one ordering tensor involving the whole triad $ \vek{n}^{\alpha}$.
These symmetries transform $\{ \mathbf{l}, \mathbf{m}, \mathbf{n} \}$ ``isotropically" amongst each other, so there is no primary axis and the three axes appear symmetrically in the order parameter tensor.  Moreover, the tetrahedral symmetries $T, T_d, T_h$ give an example of the three categories (i)-(iii) of point groups and will be discussed in Section \ref{sec:polyhedral}.
%\end{itemize}

\subsubsection{Proper point groups: chirality}

Besides the orientational order parameters, the proper point group symmetries 
$\{C_n, D_n, T, O, I \}$ are chiral and have an additional chiral order parameter. The simplest chiral order parameter is just the pseudoscalar handedness or chirality $\sigma$ of the triad defined in Eq. \eqref{eq:sigma}.
By definition, proper point groups do not possess any inversions or reflections and therefore cannot change the chirality or handedness of the triad.

\subsection{Ordering tensors and the group hierarchy} \label{sec:op_group}

As one may have already noticed from the above discussion and Table \ref{table:ops}, although a symmetry can be uniquely defined by the collection of order parameter tensors $\mathbb{O}^G$ and the handedness $\sigma$, owing to the group structure discussed above, many orientational ordering tensors are shared by different symmetries. We will now clarify this by discussing their group structures.

Firstly, the primary ordering tensor $\mathbb{A}^G[\vek{n}]$ for $C_n$ and $C_{nv}$ groups is just the order parameter tensor of the $C_{\infty}$ and $C_{\infty v}$ groups, $\mathbb{A}^{C_n}[\vek{n}] = \mathbb{A}^{C_v}[\vek{n}] = \mathbb{O}^{C_{\infty}}[\vek{n}] = \mathbb{O}^{C_{\infty v}}[\vek{n}]$.
This is due to the simple fact that $C_n$ and $C_{nv}$ groups do not transform $\mathbf{n}$, hence they differ from $C_{\infty}$ and $C_{\infty v}$  only by their in-plane structure related to $\mathbb{B}^G[\vek{l},\vek{n}]$. 
Similarly, the groups $\{ S_{2n}, C_{nh}, D_{n}, D_{nh}, D_{nd}\}$ have the same effect on $\mathbf{n}$, $\mathbf{n} \rightarrow -\mathbf{n}$. Therefore, neglecting the $\mathbf{l}$ and $\mathbf{m}$ components, these symmetries lead to the same primary ordering tensor $\mathbb{A}[\vek{n}]=\mathbb{Q}[\vek{n}]$, the uniaxial director.

Moreover, the groups $\{C_n, C_{nh}, C_{nv}, D_{n}, D_{nh} \}$ are closely related in terms of symmetries. 
$C_n$ and $C_{nh} = C_{n} \times \{ \id, \sigma_h\}$ only differ by a reflection $\sigma_h: \vek{n}\to -\vek{n}$ in the horizontal mirror $(\vek{l},\vek{m})$-plane perpendicular to $\mathbf{n}$. Thus $C_n$ and $C_{nh}$ have the same in-plane structure leading to the same secondary order parameter $\mathbb{B}^{C_n}[\vek{l},\vek{m}]$.
For the point groups $\{C_{nv}, D_{n}, D_{nh} \}$,  we have $D_{nh} = D_n \times \{ \id, \sigma_h\}$ and $C_{nv}$ and $D_{n}$ can be represented as semi-direct products $C_{nv} = C_n \rtimes \{ \id, \sigma_v\}$ and $D_{n} = C_n \rtimes \{ \id, c_2(\mathbf{l})\}$, where $\sigma_v$ is a reflection $(\mathbf{l}$,$\mathbf{n})$-plane and $c_2(\mathbf{l})$ is a two-fold rotation around the axis $\mathbf{l}$,
\begin{align}
\sigma_v = \sigma_{\mathbf{ln}} = 
\left(
\begin{array}{ccc}
 1  & 0 & 0 \\
 0 & -1 & 0 \\
 0 & 0 & 1
\end{array}
\right),
\quad
c_2(\mathbf{l}) = 
\left(
\begin{array}{ccc}
 1  & 0 & 0 \\
 0 & -1 & 0 \\
 0 & 0 & -1
\end{array}
\right).
\end{align} 
It is immediately clear that, $\sigma_v$ and $c_2(\mathbf{l})$ have the same action on the $\mathbf{l}$ and $\mathbf{m}$ components.
Therefore, $\{C_{nv}, D_{n}, D_{nh} \}$ also have the same in-plane order parameter $\mathbb{B}[\vek{l},\vek{m}]$.

The common structures of the finite axial groups have a direct implication on the associated phase transitions.
For a phase with the symmetry of a finite axial group, it is in principle possible to disorder the primary and secondary order separately before the transition to the isotropic phase.
If we first disorder the secondary order in a plane, the following sequences of phase transitions can happen 
\begin{align} \label{eq:aixal_transition}
 & C_n, C_{nv} \rightarrow C_{\infty v} \rightarrow O(3), \nonumber \\
 & S_{2n}, C_{nh}, D_{n}, D_{nh}, D_{nd} \rightarrow D_{\infty h} \rightarrow O(3),
\end{align}
related to the restoration of the in-plane $O(2)$ symmetry followed by disordering of  order along the principal axis $\vek{n}$. These axial transitions generalize the biaxial-uniaxial-isotropic liquid transition of biaxial liquid crystals \cite{Freiser70, BiaxialBook2015}.
We have numerically verified the transition sequences in Eq. \eqref{eq:aixal_transition} for a large number of symmetries and will present the detailed analysis and their phase diagrams in a separate work \cite{LiuEtAl16}.

Finally, in the case of the poyhedral groups,  $T_h = T \times \{\id,-\id\}$, $O_h = O \times \{\id, -\id\} $ and $I_h = I \times \{\id, -\id\}$ are generated from the proper subgroups $T$, $O$ and $I$ by adding the inversion $-\id$, according to item (ii) in Sec. \ref{sec:class}.
Since the ordering tensors of $I$ and $O$ in Table \ref{table:ops} are of even rank, this difference is not reflected directly in the orientational order parameters.
There exist higher order invariant tensors that can distinguish $O$ ($I$) from $O_h$ ($I_h$), nonetheless one needs to consider at least a rank-$5$ (rank-$7$) tensors and it is therefore more convenient to distinguish them by the chirality $\sigma$.

Improper groups possessing only reflections but not the inversions $-\id$ (item (iii) in Sec. \ref{sec:class}) have non-vanishing odd-rank order parameters in general. In these order parameters, terms related with right- and left-handed triads appear symmetrically, making the order parameter invariant under certain improper reflections but not inversions. 
This will be reflected in the structure of the associated order parameters. For instance, as can be seen from Table \ref{table:ops}, the order parameter for the tetrahedral-$T_d$ group, $\mathbb{O}^{T_d}$ consists of a left- and right-handed copy of that of the tetrahedral-$T$ group (see Section \ref{sec:polyhedral} for more details).
%For the tetrahedral symmetries, we have the interesting case of all three cases (i)-(iii) appearing in the groups $T,T_d, T_h$.  $T_d$ is the full symmetry group of a regular tetrahedron that does not contain the inversion, $T$ its proper subgroup and $T_h = T\times \{\id, -\id\}$, as we mentioned above. The $T_d$ order parameter is  \cite{Fel1995}

\section{Gauge theory description of generalized nematic phases} \label{sec:gaugetheory}
Let us now establish the relation with a recently introduced gauge theoretical description of the $G$-point group symmetric degrees of freedom, providing the necessary apparatus to describe generalized three-dimensional orientational order \cite{LiuEtAl2015b} and the derivation of the order parameter tensors. 

The goal is to construct a coarse-grained order parameter theory with local point group symmetries. As in the previous Section, the well-understood uniaxial nematic with $D_{\infty h}$-symmetry may be perceived as the guiding example, maximally simplified by the Abelian $\integers_2$ nature of the underlying symmetry acting on the order parameter, which should be generalized to cope with arbitrary point group symmetries. In the uniaxial case, the necessary head-to-tail symmetry that turns the coarse-grained local order parameter vector $\vek{n}_i$ into a director can be simply accomplished in a $\mathbb{Z}_{2}$ lattice gauge theoretical setting \cite{LammertRoksharToner93, LammertRoksharToner95}, see also \cite{Mettout2006}. That is, starting from an auxiliary cubic lattice regulating the short-distance cut-off of the theory, one can simply describe the coarse grained order parameter theory in terms of the $O(3)$-vectors $\vek{n}_i$ coupled to $\integers_2$ gauge fields. This theory has gauge variables $\sigma_{ij}=\pm 1$ living on the bonds $\corr{ij}$ of the lattice, that interact via a plaquette term $-K \sum_{^{l}_{i}\Box_{j}^{k}} \sigma_{ij}\sigma_{jk}\sigma_{kl}\sigma_{li} $ thereby defining Wegner's Ising gauge theory \cite{Wegner71, Kogut79}. The gauge fields are minimally coupled to the $O(3)$ vectors $\vek{n}_i$  living on the sites of the lattice via a ``Higgs" term $-J \sum_{\langle ij \rangle} \sigma_{ij} \vek{n}_i \cdot \vek{n}_j$ \cite{FradkinShenker79} favoring nematic alignment.  The theory has the local gauge symmetries where $\vek{n}_i \to -\vek{n}_i$ and $\sigma_{ij} \to -\sigma_{ij}$ simultaneously. The term with only gauge fields can be understood as encoding the $\pi$-disclinations of the uniaxial nematics. Last but not the least, integrating over the $\integers_2$ gauge fields leads directly to the effective biaxial Hamiltonian $H[\vek{n}]\sim \sum_{ij} \mathrm{Tr}~\mathbb{Q}_i \cdot \mathbb{Q}_j$ in terms of the appropriate order parameter \cite{LammertRoksharToner93}.

The above Ising lattice gauge theory is  enough to elucidate the nature of gauge theories in general \cite{FradkinShenker79}. For large $J$ the matter and gauge fields are ordered by the Higgs mechanism. In addition, the coupling $K$ controls the gauge fields and for small $K,J$ the gauge fields are confined, while  for large $K$ and small $J$ the system is in a deconfining phase with topological gauge fluxes as excitations. Such ``topological nematic phases" \cite{LammertRoksharToner95} have been identified only in strongly interacting electron systems \cite{SenthilFisher2000, SenthilFisher2001, NussinovZaanen2002, PodolskyDemler2005, GroverSenthil2011, XuLudwig2012, LiuEtAl15, Beekman2016}, while  they appear rather unphysical in the classical liquid crystal setting, where the regime of interest is the strong gauge coupling $K \rightarrow 0$ limit. Here one finds the fully ordered ``Higgs phase" and a fully disordered confining phase, which encode for the uniaxial nematic phase and the isotropic liquid, respectively. 

\subsection{General lattice gauge theory model}
As emphasized above,  the $D_{\infty h}$-symmetry is special due to the $O(2)$ symmetry in the plane perpendicular to the ordering axis. In general cases one has to revert to coarse-grained $O(3)$ {\it rotors } parametrized in terms of a \emph{local} orthornormal triad as
\begin{align}
R_i = \big( \mathbf{l}_i \quad \mathbf{m}_i \quad \mathbf{n}_i \big)^{\mathrm{T}}
\end{align}
according to some fixed coordinate basis $\vek{e}_a$, instead of a single vector $\vek{n}_i$. In the following, we denote the components of $R_i$ as $(R_i)^{\alpha}_{a} = (\vek{n}^{\alpha}_i)_a$, where greek indices are in the local orthonormal basis $\vek{n}_i^{\alpha} = \{\vek{l}_i,\vek{m}_i,\vek{n}_i\}_{\alpha =1,2,3}$ of the triad and latin indices in the fixed coordinate basis. The order parameter triad $\vek{n}_i^\alpha$ defines the \emph{local} orientation of a $G$-symmetric ``mesogen" and therefore is subjected to the \emph{local} identifications
\begin{align}
R_i \simeq \Lambda_i R_i, \quad \vek{n}^\alpha \simeq \Lambda_i^{\alpha\beta}\vek{n}_i^\beta, \quad \Lambda_i\in G.
\end{align}
Again, the goal is to construct a lattice model that gives rise to a Higgs phase, which now describes generalized nematic order. Specifically, the Hamiltonian of the model is defined as \cite{LiuEtAl2015b}
\begin{align}
H = -\sum_{\corr{ij}} \mathrm{Tr}~[R_i^\mathrm{T} \mathbb{J} U_{ij} R_j]  - \sum_{\Box} \sum_{\mathcal{C}} K_{\mathcal{C}}\delta_{\mathcal{C}}(U_{\Box}) \mathrm{Tr}~[U_{\Box}] \label{eq:gauge theory}
\end{align}
on an auxiliary cubic lattice $\integers^3$. Here the first term is a Higgs term for the matter $\{R_i\}$ and gauge fields $\{U_{ij}\}$ and in the second term 
\begin{align}
U_{\Box} = \prod^{\rightarrow}_{\corr{ij} \in \partial \Box} U_{ij}
\end{align}
is the counterclockwise product of the gauge fields around a plaquette $\Box$ of the lattice, defining a discrete gauge field strength. Moreover, the gauge symmetries act on the rotors $R_i \in O(3)$ as
\begin{align}
R_i \to \Lambda_i R_i ,\quad U_{ij} \to \Lambda_i U_{ij} \Lambda_j^\mathrm{T}, \quad U_{ij}, \Lambda_i\in G. \label{eq:gauge symmetry}
\end{align}
Under a gauge transformation $\Lambda_i$, the plaquette term transforms as as $U_\square \to \Lambda_i U_\square \Lambda_i^{-1}$ and hence the gauge field strength $U_\Box$ is defined only up to conjugation and $K_{\mathcal{C}}\delta_\mathcal{C}(\bullet)$ is a function on the conjugacy classes $\mathcal{C}$ of the group $G$. In  addition, the model is symmetric under global $O(3)$ rotations
\begin{align}
R_i \to R_i \Omega^\mathrm{T}, \Omega \in O(3)
\end{align}
which is the global symmetry that the nematic state breaks spontaneously.

The generalization from to the Ising gauge theory relevant in the context of $D_{\infty h}$ nematics is obvious. The Higgs term models the orientational interaction of the matter fields $R_i$, where 
$\mathbb{J}$ is now a $G$ symmetric coupling matrix,
\begin{equation}
 \Lambda \mathbb{J} \Lambda^{\mathrm{T}} = \mathbb{J},\qquad  \forall \Lambda \in G,
\end{equation}
defining the couplings between the triads, while the second term is a defect suppression term. Accordingly, $K_{\mathcal{C}}$ describes the core energy of a defect corresponding to deficit angle $U_{\square}\in G$ defined up to the conjugacy classes $\mathcal{C}$ of the group $G$. These gauge defects do not directly classify topological defects in nematics, but are nonetheless closely related via the so-called Volterra construction \cite{Friedel64, Kleinert89b, KlemanFriedel08}.
The nematic defects are usually classified topologically in terms of the homotopy groups of the manifold $O(3)/G$ \cite{Mermin1979, Michel1980} which is the order parameter space of the $G$-nematic and as well the low-energy manifold of the model Eq.\eqref{eq:gauge theory} in the Higgs phase. Disordered configurations can be suppressed by assigning a finite core-energy to the gauge defects that proliferate at the phase transition.
Thus, $K_{\mathcal{C}}$ can effectively be regarded as tuning the fugacity of the nematic defects. However, the interesting ``deconfined" regime at large $K_{\mathcal{C}}$  will be left for future work  and we stress that in the present  context of deriving the order parameters using the gauge symmetries, we can focus on the gauge theory in the limit $K_\mathcal{C} = 0$.

In the usual way, the Hamiltonian defines the orientational probability density as $\corr{\mathbb{O}[R_i]} = \frac{1}{Z} \sum_{\{R_i\}, \{U_{ij}\}} \mathbb{O}[R_i] e^{-\beta H}$, however due to the gauge symmetries only very particular quantities $\mathbb{O}[R_i]$ are non-trivial, as we next discuss.

\subsection{Gauge theory, generalized nematics and order parameters}
With the gauge theory at hand, we need to make the connection to generalized nematics in the ordered Higgs phase where the $O(3)$ symmetry spontaneously breaks. We will first discuss how it is automatically guaranteed that the ordering is of nematic nature and then outline the construction of the $G$-symmetric order parameter tensor. 

Higher rank order parameter tensors like those in Table \ref{table:ops} arise in fact naturally  from gauge invariant tensors of the matter fields $R_i$ in our gauge model Eq. \eqref{eq:gauge theory}. That is, due to Elitzur's theorem \cite{Elitzur75}, all gauge non-invariant tensors vanish. 
Therefore, on each site, instead of the matter fields $R_i$ we have the local tensors $\mathbb{O}_i = R_i \otimes R_i \otimes \cdots \otimes R_i$ carrying upper and lower indices $\alpha, \beta, \dots, \gamma$ and $a, b ,\dots, c$ giving rise to physical degrees of freedom. Each choice of the triad indices $\alpha, \beta, \dots \gamma$ leads to a tensor
\begin{align}
(\mathbb{O}^{\alpha \beta \cdots \gamma}_i)_{a b \cdots c} = (R_i )^{\alpha}_a\otimes (R_i)^{\beta}_b \otimes \cdots \otimes (R_i)^\gamma_c , \label{eq:tensor basis}
\end{align}
with components denoted by the Latin indices $a,b,\dots, c$ on which the global $O(3)$ rotations act. Such ``bi-tensors" $\mathbb{O}_i$ are sometimes referred to as supertensors in the soft-matter literature \cite{DeGennesProst95, Mettout2006, Rosso2007}. Due to the $O(3)$ constraints, i.e. the fact that $\vek{n}_i^{\alpha}$ is a local orthonormal triad, not all tensors $\mathbb{O}^{\lambda}_i$ are independent: the simplest such tensor relations are given by $\vek{l}\otimes \vek{l} + \vek{m}\otimes \vek{m} + \vek{n} \otimes \vek{n} = \id$ and Eq. \eqref{eq:sigma} but we will come back to these later in Section \ref{sec:examples}. 

The gauge symmetries act on the indices $\alpha, \beta, \dots, \gamma$ and transform these tensors onto linear combinations under the gauge transformations. Formally we can write the gauge invariant tensors as linear combinations 
\begin{align}
\mathbb{O}^{\boldsymbol{\lambda}}_i = \sum_{\{\lambda\}} c_{\lambda}\mathbb{O}^{\lambda}_i,
\end{align}
where each multi-index $\lambda = \alpha\beta \dots \gamma$ determines a tensor in  Eq. \eqref{eq:tensor basis} with the coeffiecients  $c_{\lambda} = c_{\alpha\beta \cdots \gamma}$. The full tensor $\mathbb{O}_i^{\boldsymbol{\lambda}}$ is uniquely determined by $\boldsymbol{\lambda} = \sum_{\{\lambda\}} c_\lambda \vek{e}_{\lambda}$ (vector of tensor components). 
The components transform as $R_{\mathrm{L}}(\Lambda) \cdot c_{\lambda} = (\Lambda^T)^{\alpha' \alpha}(\Lambda^T)^{\beta' \beta} \cdots (\Lambda^T)^{\gamma \gamma'} c_{\alpha'\beta' \cdots \gamma'}$ under the gauge transformations (local rotations of the triad). By Elitzur's theorem, we thus have 
\begin{align}
\corr{\mathbb{O}_i^{\boldsymbol{\lambda}}} = 0, \quad \textrm{ if } \corr{\mathbb{O}_i^{\boldsymbol{\lambda}}} \neq R_{\mathrm{L}}(\Lambda_i) \cdot \corr{\mathbb{O}_i^{\boldsymbol{\lambda}}} := \corr{\mathbb{O}_i^{R_{\mathrm{L}}(\Lambda_i) \cdot \boldsymbol{\lambda}}} %\neq \corr{\mathbb{O^{\boldsymbol{\lambda}}}},
\end{align}
i.e. if $\boldsymbol{\lambda}$ is not a gauge invariant combination $R_{\mathrm L}(\Lambda)\cdot\boldsymbol{\lambda} = \boldsymbol{\lambda}$, specifying a gauge singlet tensor. Each point group $G\subset O(3)$ has a set of minimal linearly independent gauge invariant tensors $\mathbb{O}^{\boldsymbol{\lambda}}$, see Table \ref{table:ops} with the identification $\vek{n}_i^{\alpha} \simeq \{\vek{l},\vek{m}, \vek{n}\}$. Each tensor $\mathbb{O}^{\boldsymbol{\lambda}}_i$ has the rank $\vert \boldsymbol{\lambda} \vert = \dim~ \lambda = \dim (\alpha,\beta,\dots,\gamma)$ which is related to elements of finite order in the point group; if there are elements of order $n$, then the rank of the invariant tensors is at least $n$. On the other hand, Ref. \onlinecite{HajiAkbariGlotzer15} provides rigorous upper bounds on the rank of the order parameter for subgroups of $SO(3)$. All invariant combinations can be constructed as tensor products of the invariant tensors, similarly as in the case of point-group invariant polynomials (i.e. scalars) \cite{JaricSharp1984, Michel2001}.

Up to this point, we have determined that the gauge model Eq. \eqref{eq:gauge theory} gives automatically non-trivial higher rank tensors that describe long-range order associated with the spontaneous breaking of the $O(3)$ symmetry, while the ``usual" orientational order $\corr{R_i} \neq 0$ is forbidden. Therefore, the physical orientational correlations are in the tensors $\mathbb{O}^{\boldsymbol{\lambda}}[R_i]$ and the orientational probability density is given terms of the Hamiltonian as $\rho(\{R_i\}) \sim \frac{1}{Z} \sum_{\{U_{ij}\}}e^{-\beta H[\{R_i\}, \{U_{ij}\}]}$.

We still have to determine the orientational order carried by the higher rank tensors $\corr{\mathbb{O}_i^{\boldsymbol{\lambda}}}$ (see Section \ref{sec:measurement} for more details) that results from the Hamiltonian. 
The global symmetry of the nematic phase is given by a point group $G'$ in the basis $\vek{e}_a = \{\vek{e}_1,\vek{e}_2,\vek{e}_3\}$ in the sense that for $\Omega \in G'\subset O(3)$, the rotations $\Omega$ act on the tensors $\corr{\mathbb{O}_i^{\boldsymbol{\lambda}}}\sim \mathbb{O}[\vek{l},\vek{m},\vek{n}]$ in the basis $\vek{e}_a \otimes \vek{e}_{b} \cdots \otimes \vek{e}_c$ as
\begin{align}
(R_{\mathrm R}(\Omega) \cdot \corr{\mathbb{O}^{\boldsymbol{\lambda}}})_{a b \cdots c} = \Omega^{a a'}\Omega^{ b b'}\cdots \Omega^{c c'} (\corr{\mathbb{O}^{\boldsymbol{\lambda}}})_{a' b' \cdots c'},
\end{align}
as expected. We emphasize this difference with the labels $R_{\rm L,R}$ (for left- and right actions) on the rotations acting on the two distinct indices of the ``supertensors" $\mathbb{O}$. The essence is that the problem of classifying the order parameters and gauge invariant tensors is the equivalent problem of classifying point group invariant tensors in the global and local bases $\vek{e}_a$ and $\vek{n}_i^{\alpha}$, respectively. Each tensor $\mathbb{O}^{\boldsymbol{\lambda}}[\vek{l},\vek{m},\vek{n}]$ is classified by the tensor rank $\vert \boldsymbol{\lambda} \vert$, the tensor symmetries of the indices $a, b, \dots, c$, and can be decomposed in to the irreducible representations of $SO(3)$ under rotations in the usual way.

\subsection{Order parameters and the high-temperature expansion of the gauge theory} \label{sec:deriving_ops}

As we discussed, the principle underlying the derivation of the order parameters is the fundamental gauge theoretical result:  all physical observables have to be gauge invariant, since gauge symmetries cannot break spontaneously \cite{Elitzur75}. By construction, the model Eq. \eqref{eq:gauge theory} embodies the symmetry of the order parameter tensors by the gauge symmetry. Therefore, if we integrate out the gauge fields in the Hamiltonian, the terms that survive are gauge invariant local combinations of the matter fields,  corresponding to the order parameter tensors.
This can be most easily accomplished in the strong coupling limit of the gauge theory $K_{\mathcal{C}} = 0$, where the gauge fields have no independent dynamics. The result is essentially the effective Hamiltonian of the orientational probability density $\rho(\{R_i\}) \sim \frac{1}{Z} \sum_{\{U_{ij}\}}e^{-\beta H[\{R_i\}, \{U_{ij}\}]}$ but in order to find the order parameter tensors we do not need the effective Hamiltonian in closed form and utilize the high-temperature expansion for small $\beta$. The couplings $\mathbb{J}$ do not affect the general form of the expansion and we set them to be isotropic $\mathbb{J} = J\id$ for simplicity and measure the temperature in the units $\beta J \equiv \beta$.

The partition function of the model Eq.\eqref{eq:gauge theory} is defined in the usual way
\begin{align}
Z &= \sum_{ \{R_i \}} \sum_{ \{U_{ij}\} } e^{-\beta H[R_i, U_{ij}]} \nonumber \\
	 &=  \sum_{ \{\tilde{R}_i \}} \sum_{ \{\sigma_{i} \}} \sum_{ \{U_{ij}\} } e^{-\beta H[\tilde{R}_i, \sigma_{i}, U_{ij}]},
\end{align}
where the summations are naturally discrete over the lattice and discrete or continuos over the groups $G$ and $O(3)$. In the second line we made the handedness field explicit by using Eq. \eqref{eq:sigma}, $R_i = \sigma_i \widetilde{R}_i$.
In order to integrate over the gauge fields, the partition function is Taylor expanded in the high temperature limit $\beta \ll 1$,
\begin{align} \label{eq:k0_expansion}
%Z = \sum_{ \{\tilde{R}_i \}} \sum_{ \{\sigma_{i} \}} \sum_{ \{U_{ij}\} } \left(1 - \beta H + \frac{1}{2!} (-\beta H)^2 + \dots + \frac{1}{n!}(-\beta H)^n+ \ \right).
Z = \sum_{ \{\tilde{R}_i \}} \sum_{ \{\sigma_{i} \}} \sum_{ \{U_{ij}\} } \prod_{\corr{ij}}\sum_{n=0}^{\infty} \frac{1}{n!}(-\beta H_{ij})^n. 
\end{align}
The integration over the gauge fields can be explicitly performed on the lattice order by order in the expansion. By construction, the terms appearing must be local terms that are composed of contractions of gauge invariant tensors. The result is therefore an expression starting with contractions $\sim \mathrm{Tr}~[\mathbb{O}^G_i\cdot \mathbb{O}^G_j]$ coming from the lowest order non-zero terms $n_{\rm min} \sim \textrm{rank }\mathbb{O}^{G}$ in the expansions. In other words, the lowest order non-trivial terms are composed of the lowest order invariant tensors that can be found from Table \ref{table:ops}. We emphasize that  by construction these tensors are the minimal and simplest possible set of invariant tensors allowed by the symmetries.

\subsection{Determining the symmetry of a phase with a high-rank order parameter} \label{sec:measurement}

%Finally, before discussing the order parameters in Table \ref{table:ops} in more detail and how they are obtained using the gauge theory, let us briefly discuss how they were used in simulations of the lattice gauge theory in our previous paper \cite{LiuEtAl2015b}.

Finally, we come back to the issue of determining the symmetry of a nematic phase. The ordering tensors we shown in Table \ref{table:ops} generalize the local director tensor $\mathbb{Q}_{ab}$ for uniaxial nematics.
The macroscopic order parameters are defined as coarse grained averages over the system
\begin{align}
\langle \mathbb{O}^{G} \rangle = \frac{1}{V} \sum_i \corr{\mathbb{O}^{G}_i},
\end{align}
where $V$ denotes the spatial averaging volume. To verify the symmetry of a phase, one need in principle consider all independent entries of the order parameter tensor. This is in general quite involved since the number of the entries grows exponentially with the rank of the tensor.

However, for interactions favoring homogenous a nematic order, such as the interaction in the gauge model Eq. \eqref{eq:gauge theory}, the symmetry of the phase can be revealed by the scalar two point functions in the limit of large separation.
Since $\langle \mathbb{O}^{G}_i \rangle$ will develop a finite value in the ordered phase, at long distances the scalar two point function of the order parameter tensor behaves as
\begin{align}
& \lim_{ \lvert i-j \rvert \rightarrow \infty} 
\langle (\mathbb{O}^G_{i})_{abc\dots} (\mathbb{O}^G_{j})_{abc\dots} \rangle   \nonumber\\
&  = 
		\begin{cases}
				\ \mathrm{Tr}~ \langle \mathbb{O}^G_{i} \rangle^2 >0 \ &\text{nematic} \\
				\ 0 \ &\text{otherwise}. 
		\end{cases} 
\end{align}
The contractions in $\rm{Tr}~(\bullet)$ are determined up to the tensor symmetries of the order parameter. This allows us to define a strength for the ordering tensors,
\begin{align}
q = \sqrt{\langle (\mathbb{O}^{G}_{i})_{abc\dots} \rangle^2},
\end{align}
and the symmetry of the phase can be defined by the lowest order tensor and ``smallest" group $G$ with $q \neq 0$.
Accordingly, the phase transition(s) associated with $\langle \mathbb{O}^{G}_i \rangle $ can be identified from the susceptibility $\chi(q)$ of the ordering strength,
\begin{align} \label{eq:susceptibility}
\chi(q) =  \beta V \big(\corr{q^2} - \corr{q}^2 \big). 
\end{align}
We have previously numerically computed $q$ and $\chi(q)$ in the model Eq. \eqref{eq:gauge theory} for large number of point group symmetries \cite{LiuEtAl2015b}.
Our simulations showed that $\chi(q)$ will exhibit a clear peak at the temperature where the heat capacity peaks, indicating that $q$ in combination of simple symmetry arguments  is indeed sufficient to determine the symmetry of a nematic phase with homogeneous distribution of order parameters.

Lastly, we note that, when non-homogeneous distributions of order parameters are preferred, the symmetry of a state can be compatible but not identical to $G$, as also discussed e.g. in Ref. \cite{Mettout2006}. In these cases, a non-zero $q$ is not sufficient to identify the symmetry of the state, and one in principle need consider all components of $\corr{\mathbb{O}_i^{G}}$.
However, the symmetry of a phase may be also determined by the ``eigenvalues'' and the distribution of non-zero entries of  $\corr{\mathbb{O}_i^{G}}$ \cite{ZhengPalffyMuhoray07}.
Studies with this regard so far mostly concentrate on the rank-$2$ $D_{\infty h}$ and $D_{2h}$ ordering tensors \cite{Turzi2016, ChillingworthTurzi2015, Turzi2011, Rosso2007}, it would be interesting to consider the ordering of the tensors in Table \ref{table:ops} in full generality without assumptions on microscopic configurations of a particular model.

\section{Examples and discussion} \label{sec:examples}

Let us finally present how we arrived to Table \ref{table:ops}. For all finite and discrete point groups, we can integrate over the gauge fields in the expansion \eqref{eq:k0_expansion}. For the continuous axial groups, we can do the integrations in closed form. The results are by construction composed of local contractions of the simplest gauge invariant tensors allowed by the symmetries, i.e. the tensors in Table \ref{table:ops}.

\subsection{Continuous axial groups: unixial nematics}

The integration over the gauge groups $\{C_{\infty}, C_{\infty v}, C_{\infty h}, D_{\infty}, D_{\infty h}\}$ can be done leading to the familiar results. We will use the $D_{\infty}$-uniaxial nematic as an example of the general procedure of deriving uniaxial nematic order parameters, the others being similar. The key point is the elimination of the triad vectors $\vek{l},\vek{m}$ in the plane where the $SO(2)$-symmetry acts from the Hamiltonian upon integrating  out  the $SO(2)$-gauge fields, since there can be no gauge invariant combinations of these components.

The gauge fields $U_{ij} \in D_{\infty}$ can be generated by the transformations $\{c_{\theta}(\mathbf{n}), c_2(\mathbf{m}) \}$, where
\begin{align} \label{eq:op_generator_dinfh}
&c_{\theta}(\mathbf{n}) = 
\left(
\begin{array}{ccc}
 \cos \theta & -\sin \theta & 0 \\
 \sin \theta & \cos \theta & 0 \\
 0 & 0 & 1
\end{array}
\right),
c_2(\mathbf{m}) = 
\left(
\begin{array}{ccc}
 -1  & 0 & 0 \\
 0 & 1 & 0 \\
 0 & 0 & -1
\end{array}
\right)
\end{align}
are a rotation about $\mathbf{n}$ by an angle $\theta \in [0, 2\pi)$ and a $\pi$-rotation about $\mathbf{m}$, respectively. We note that the ``usual" uniaxial symmetry is given by $D_{\infty h} = D_{\infty} \times \{\id, -\id\}$ and follows with minimal modifications. 
We focus on the terms in the $(\vek{l}, \vek{m})$-plane and parametrize the gauge transformation as 
\begin{align}
U_{ij} =\begin{matriisi} \sigma_{11} \cos\theta_{ij} & \sigma_{12}\sin \theta_{ij} & \\ -\sigma_{21} \sin \theta_{ij} & \sigma_{22} \cos \theta_{ij} & \\ & & \sigma_{33} \end{matriisi} \in D_{\infty},
\end{align}
where $\theta_{ij}\in [0,2\pi)$ parametrizes the $C_{\infty}$ rotation and the constrained signs $\sigma_{\alpha\beta} = \pm 1$ are determined by the presence of the $\pi$-rotation in the orthogonal $(\vek{l},\vek{n})$-plane. 
This gives from Eq. \eqref{eq:gauge theory}, with $\mathbb{J} = J \id$ and $\beta\equiv \beta J$ as before,
\begin{align}
&\beta H[\vek{l},\vek{m},\vek{n}, \theta, \sigma_{\alpha\beta}] \nonumber \\
&= \sum_{\corr{ij}} \beta\bigg[\cos \theta_{ij} \big( \sigma_{11} \vek{l}_i \cdot \vek{l}_j + \sigma_{22} \vek{m}_i\cdot\vek{m}_j \big) \nonumber \\
& \quad + \sin \theta_{ij} \big(\sigma_{12} \vek{l}_i\cdot \vek{m}_j- \sigma_{21} \vek{m}_i\cdot\vek{l}_j\big)  + \vek{n}_i U_{ij,33} \cdot\vek{n}_j\bigg].
\end{align}
Now we proceed to integrate over the $SO(2)$ angle $\theta_{ij}$
\begin{align}
&e^{-\beta H_{\textrm{eff}}[\vek{l}_i, \vek{l}_j,\vek{m}_i,\vek{m}_j, \sigma_{\alpha\beta}]} \nonumber \\
&= \prod_{\corr{ij}} \f{1}{2\pi}\int^{2\pi}_{0}\de \theta_{ij}~ e^{-H[\vek{l}_i, \vek{l}_j,\vek{m}_i,\vek{m}_j,\theta_{ij}, \sigma_{\alpha\beta}]} \nonumber \\
&= \prod_{\corr{ij}} I_0(J_1 \sqrt{A^2_{ij} + B^2_{ij}} ).
\end{align}
where $I_0(z)$ is a Bessel function of the first kind with the argument
\begin{align*}
&A^2_{ij} +B^2_{ij}  \nonumber \\
&= \big[ \sigma_{11} \vek{l}(i) \cdot \vek{l}(j) + \sigma_{22} \vek{m}(i)\cdot\vek{m}(j) \big]^2  \nonumber \\
& \quad + \big[\sigma_{12} \vek{l}(i)\cdot \vek{m}(j)- \sigma_{21} \vek{m}(i)\cdot\vek{l}(j)\big]^2  \nonumber \\
&= (\vek{l}_i\cdot\vek{l}_j)^2 + (\vek{m}_i\cdot\vek{m}_j)^2 + (\vek{m}_i\cdot\vek{l}_j)^2 \nonumber \\
&\quad +(\vek{l}_i\cdot\vek{m}_j)^2 + 2\sigma_{11}\sigma_{22} (\vek{m}_i\cdot\vek{m}_j)(\vek{l}_i\cdot\vek{l}_j) \nonumber \\
& \quad - 2\sigma_{12}\sigma_{21}(\vek{l}_i\cdot\vek{m}_j)(\vek{m}_i\cdot\vek{l}_j).
\end{align*}
Now, since $\det_{2\times 2} U_{ij} = \sigma_{11}\sigma_{22}\cos^2 \theta_{ij}+\sigma_{12}\sigma_{21}\sin^2\theta_{ij} = \pm 1 = \det U_{ij}\times U_{33,ij}$, we can simplify 
\begin{align*}
&A^2_{ij} +B^2_{ij} \nonumber \\
&=  (\vek{l}_i\cdot\vek{l}_j)^2 + (\vek{m}_i\cdot\vek{m}_j)^2 + (\vek{m}_i\cdot\vek{l}_j)^2 +(\vek{l}_i\cdot\vek{m}_j)^2 \nonumber \\
&\quad +2 \det_{2\times 2} U_{ij} \big[ (\vek{m}_i\cdot\vek{m}_j)(\vek{l}_i\cdot\vek{l}_j) - (\vek{l}_i\cdot\vek{m}_j)(\vek{m}_i\cdot\vek{l}_j)\big] \nonumber \\
&= 1+ (\vek{n}_i\cdot\vek{n}_j)^2 + 2 \det_{2\times2} U_{ij} \sigma_i \sigma_j\vek{n}_i\cdot\vek{n}_j \nonumber \\
&= (\sigma_i\sigma_j \vek{n}_i\cdot\vek{n}_j + \det_{2\times 2} U_{ij})^2,
\end{align*}
where on the second-to-last line we used the $O(3)$ relation $\vek{l}_i \times \vek{m}_i = \sigma_i \vek{n}_i$. Using $\det_{2\times 2} U_{ij} = U_{ij, 33}$ gives the result
\begin{align}
&H_{\textrm{eff}}[\vek{n}_i,U_{ij}] =  \nonumber \\
&-\sum_{\corr{ij}} \beta \vek{n}_i\cdot U_{ij,33}\vek{n}_j 
+ \log I_0\big(\beta |\sigma_i\sigma_j \vek{n}_i\cdot\vek{n}_j +U_{ij,33}|\big), 
%\\  &=& \sum_{\corr{ij}} J_3 \vek{n}_i\cdot U_{ij,33}\vek{n}_j +2 \log I_0\big(J_1|\vek{n}_i\cdot\vek{n}_j+1|\big).
\end{align}
where $U_{ij,33}= \pm 1\in\integers_2$ since for $U_{ij} \in D_{\infty}/C_{\infty} \simeq \{\id, c_2(\vek{m})\} = \integers_2$ when acting on $\vek{n}_i$. We remind that
\begin{align}
I_0(z) = \sum_{k=0}^{\infty} \frac{(z^2/4)^{k}}{(k!)^2}, %\sim \f{e^z}{\sqrt{2\pi z}}(1+\f{1}{8z}+O(1/z^2))
\end{align}
meaning that to lowest order in $\beta$, we generate the term % and $1/\beta$, we generate for $\beta\ll 1$ the terms
\begin{align}
&\delta H_{\rm{eff}}[\vek{n}_i, U_{ij, 33}] \nonumber \\
&\sim \sum_{\corr{ij}}\frac{\beta^2}{4} \bigg[1+2 \sigma_i \sigma_j \vek{n}_{i}\cdot U_{ij,33}\vek{n}_j + (\vek{n}_i\cdot\vek{n}_j)^2\bigg] + O(\beta^4) \nonumber\\ 
&\sim \sum_{\corr{ij}} \frac{\beta^2}{2} \widetilde{\vek{n}}_i U_{ij,33}\cdot \widetilde{\vek{n}}_j + \textrm{higher orders},
\end{align}
%and for $\beta\gg 1$
%\begin{align}
%\delta H_{\rm{eff}}[\vek{n}] &\sim \sum_{\corr{ij}}\beta \abs{ \sigma_i \sigma_j \vek{n}_i\cdot\vek{n}_j + U_{ij,33}} \nonumber
%\\ & - \frac{1}{2}\log 2\pi \beta\abs{ \sigma_i \sigma_j \vek{n}_i\cdot\vek{n}_j + U_{ij,33}} + O(1/\beta) \nonumber \\
%&\quad \sim \sum_{\corr{ij}}\beta\abs{ \widetilde{\vek{n}}_i \cdot \widetilde{\vek{n}}_j + U_{ij,33}} + \textrm{const}.
%\end{align}
in addition to the original Hamiltonian in terms of $\vek{n}_i$. By integrating out $U_{ij,33} \in \integers_2$ one will find that all odd powers of $\vek{n}_i\cdot \vek{n}_j$ vanish and the first non-trivial term is second order with $D_{\infty}$-invariant scalar contractions
\begin{equation}
(\tilde{\mathbf{n}}_i \cdot \tilde{\mathbf{n}}_j)^2 = (\mathbf{n}_i \cdot \mathbf{n}_j)^2 =\mathrm{Tr }[\mathbb{Q}_i \cdot \mathbb{Q}_j] + \mathrm{const} \label{eq:op_polynomial_dinfh}
\end{equation}
due to the fact that a pseudovector and a vector are indistinguishable for even powers.
At the same time, this is the minimal $D_{\infty h}$-invariant tensor contraction $\mathrm{Tr }[\mathbb{Q}_i \cdot \mathbb{Q}_j]$. Higher order terms in Eq.\eqref{eq:k0_expansion} are high order even functions such as $[(\mathbf{n}_i \cdot \mathbf{n}_j)^2]^2$, and $[(\mathbf{n}_i \cdot \mathbf{n}_j)^2]^3$  etc. that can be neglected as irrelevant. Note however, that the full expansion \eqref{eq:k0_expansion} for $D_{\infty}$ contains odd powers of $\beta$ with terms of the form $\beta^3 \sigma_i \sigma_j [ (\vek{n}_i \cdot \vek{n}_j)^2 +\cdots]$ that feature the chiral order parameter $\sigma_i$. These chiral terms vanish identically for the case $D_{\infty h}$ when summing over the gauge fields $U_{ij} = \{\id, -\id\}$ in $D_{\infty h} = D_{\infty}\times \{\id, -\id\}$.
%\textcolor{red}{... end\{appendix\} ???}

\subsection{Biaxial nematics}
The $D_{\infty}$- and $D_{\infty h}$-uniaxial nematics we just discussed is a well-known and relatively simple case in the generalized nematic family. Since the symmetries $\{ C_{\infty}, C_{\infty v}, C_{\infty h}, D_{\infty}, D_{\infty h}\}$ all contain a $SO(2)$ part in the plane perpendicular to the $\vek{n}$, the vectors $\vek{l}$ and $\vek{m}$ disappear from the order parameter, as we saw above.
For the symmetries  $\{ C_n, C_{nv}, C_{nh}, S_{2n}, D_n, D_{nh}, D_{nd}\}$ with finite $n$, however, there will be in-plane rotational symmetry breaking and we need a secondary ``biaxial" order parameter $\mathbb{B}[\vek{l},\vek{m}]$ or $\mathbb{B}[\vek{l},\vek{m}, \mathbf{n}]$ to capture these phase transitions.

\subsubsection{Biaxial order parameters and $O(3)$-constraints}
Let us first discuss second rank order parameter tensors. All traceless second order parameter tensors can be generated from the ``supertensor" \cite{DeGennesProst95, Rosso2007}
\begin{align}
\mathbb{S}^{\alpha\beta} = \vek{n}^{\alpha} \otimes \vek{n}^{\beta} -\frac{\id}{3}\delta^{\alpha\beta}.
\end{align}
However, as expected, there are only 5 independent traceless symmetric second-rank tensors, due to the relation $\mathbb{S}^{\alpha\alpha} = 0$
following from the $O(3)$ constraints
\begin{align} \label{eq:O3_relation}
\vek{l}\otimes \vek{l} + \vek{m}\otimes \vek{m}+\vek{n}\otimes \vek{n} = \id.
\end{align}
Therefore symmetric combinations of $\mathbb{S}^{\alpha\beta}$ have in total 25 non-zero elements (i.e. ``scalar" order parameters). However, these components transform under rotations of the laboratory and body axes $\vek{n}^{\alpha}$ and therefore are constrained by the symmetries. In general we can eliminate three non-zero components with a rotation to a suitable laboratory frame that diagonalizes a symmetric second rank order parameter tensor derived from $\mathbb{S}^{\alpha\beta}$. Similarly it is possible to consider such relations for all higher order ``supertensors" $\mathbb{O}^{\alpha\beta\dots\gamma}$ and decompose them into irreducible $SO(3)$-representations according to their rank.

Let us now give some examples in terms of the familiar order parameters of biaxial nematics. As can be seen from the Table \ref{table:ops}, for some axial nematics, there exist more than one biaxial order parameter $\mathbb{B}$. A well-known example is the biaxial $D_{2h}$-nematic, 
where we have the order parameters $\{ \mathbb{B}^{D_{2h}}_1,  \mathbb{B}^{D_{2h}}_2\}$,
\begin{subequations} \label{eq:op_D2h_B}
\begin{align}
\mathbb{B}^{D_{2h}}_{1} &= \mathbb{S}^{\vek{l}\vek{l}} =  \mathbf{l} \otimes \mathbf{l} - \frac{\id}{3}, \\
\mathbb{B}^{D_{2h}}_{2} &=  \mathbb{S}^{\vek{m}\vek{m}}=  \mathbf{m} \otimes \mathbf{m} - \frac{\id}{3}
\end{align}
\end{subequations}
which are both clearly invariant under $D_{2}$ generated by $\{c_2(\vek{n}),c_2(\vek{l})\}$ and as well as the inversion $-\id$. Correspondingly, when integrating over $U_{ij} \in D_{2h}$ in the expansion Eq.\eqref{eq:k0_expansion}, in the first non-trivial order one will obtain the scalar contractions 
\begin{align} \label{eq:D2h_expansion}
&\sim (\vek{l}_i\cdot\vek{l}_j)^2 + (\vek{m}_i\cdot\vek{m}_j)^2 +(\vek{n}_i\cdot\vek{n}_j)^2 \nonumber \\
&= \mathrm{Tr }[\mathbb{Q}\cdot\mathbb{Q}] + \mathrm{Tr }[\mathbb{B}^{D_{2h}}_{1}\cdot\mathbb{B}^{D_{2h}}_{1}] + \mathrm{Tr }[\mathbb{B}^{D_{2h}}_{2}\cdot\mathbb{B}^{D_{2h}}_{1}]  + \mathrm{const.}
\end{align}
 which cannot be written as a contraction a single local quantity like in Eq. \eqref{eq:op_polynomial_dinfh}. However, due to the $O(3)$-constraints Eq. \eqref{eq:O3_relation}, the commonly used $D_{2h}$ biaxial order parameter tensor $\mathbb{B}^{D_{2h}} = \mathbf{l} \otimes \mathbf{l} - \mathbf{m} \otimes \mathbf{m}$
is just the linearly independent linear combination of $\{ \mathbb{B}^{D_{2h}}_1, \mathbb{B}^{D_{2h}}_2\}$ and  Eq. \eqref{eq:D2h_expansion}
reduces to contractions of the two independent rank-$2$ tensors. In summary, the biaxial $D_{2h}$ gauge symmetry eliminates all non-diagonal elements of $\mathbb{S}^{\alpha \beta}$ and the $O(3)$ constraints give one linear relation between the order parameters. 

For other two-fold symmetries with second rank order parameter tensors, similar arguments hold. For example for $C_{2h}$ symmetry, the tensors 
\begin{subequations} \label{eq:op_C2h_B}
\begin{align}
\mathbb{B}^{C_{2h}}_{1} &= \mathbb{S}^{\vek{l}\vek{m}} =  \mathbf{l} \otimes \mathbf{m}, \\
\mathbb{B}^{C_{2h}}_{2} &=  \mathbb{S}^{\vek{m}\vek{l}}=  \mathbf{m} \otimes \mathbf{l}
\end{align}
\end{subequations}
 are invariant in addition to the $D_{2h}$ order parameters and therefore in general non-zero. 
 In terms of the irreducible representations of $SO(3)$ and the $O(3)$ constraints, their combination lead to the following order parameters
\begin{align}
\mathbb{B}^{C_{2h}}_{1} + \mathbb{B}^{C_{2h}}_{2} & = \vek{l}\otimes\vek{m} + \vek{m}\otimes \vek{l} \\
\mathbb{B}^{C_{2h}}_{1} - \mathbb{B}^{C_{2h}}_{2}  &= \vek{l}\otimes\vek{m} - \vek{m}\otimes \vek{l} \nonumber \\
&= \widehat{\epsilon} \cdot \vek{n} \simeq \sigma\vek{n} = \mathbb{A}^{C_{\infty h} [\widetilde{\mathbf{n}}]},
\end{align}
which respectively correspond to the symmetric traceless second rank tensor and the pseudovector $\widetilde{\vek{n}} = \sigma \vek{n}$ representations of $O(3)$, where $(\widehat{\epsilon} \cdot \vek{n})_{ab} = \epsilon_{abc}n_c$.
These were precesicely the rank-$2$ tensor order parameters discussed in Ref. \cite{LuckhurstEtAl11} for $C_{2h}$ symmetry.

\subsubsection{Generalized biaxial order parameters}

To show how more complicated order parameters are derived using the gauge theory, we next discuss the derivation of the secondary in-plane order parameters $\mathbb{B}^G$ of higher rank using the the order parameters of $D_{2d}$, $D_{4h}$ and $C_{6h}$ symmetries as examples.

We take $D_{2d}$ symmetry as an example of a nematic with third-rank order parameter. The $D_{2d}$ group is generated by the elements $\{ c_2 (\mathbf{n}), c_2 (\mathbf{m}), \sigma_{d} \}$, where 
\begin{align} \label{eq:D2d_generator}
c_2(\mathbf{n}) = 
\left(
\begin{array}{ccc}
 -1  & 0 & 0 \\
 0 & -1 & 0 \\
 0 & 0 & 1
\end{array}
\right),
%\sigma_{\vek{l}+\vek{m},\vek{n}} = 
\sigma_d =
\left(
\begin{array}{ccc}
 0  & -1 & 0 \\
 -1 & 0 & 0 \\
 0 & 0 & 1
\end{array}
\right)
\end{align}
are a $2$-fold rotation about $\mathbf{n}$ and a reflection in the $(\vek{l}+\mathbf{m},\mathbf{n})$ plane, respectively, and $c_2(\mathbf{m})$ is as that in Eq.\eqref{eq:op_generator_dinfh}.

These lead to a $4$-fold symmetry in the $(\mathbf{l},\mathbf{m})$-plane. To obtain the order parameter describing this symmetry breaking, we follow the same procedure discussed in the previous section, but now the gauge fields $U_{ij}$ in Eq.\eqref{eq:k0_expansion} are elements of $D_{2d}$.
Integrating over $U_{ij} \in D_{2d}$, one will find that the first non-trivial order is the second order with a term 
$(\mathbf{n}_i \cdot \mathbf{n}_j)^2$, which indicates as expected that $\mathbb{Q}[\vek{n}]$ is as well an order parameter for $D_{2d}$ nematics.
The $4$-fold rotational symmetry combined with the reflections starts showing up at the third order in Eq.\eqref{eq:k0_expansion}, where one finds the following contractions up to a constant factor
\begin{align} \label{eq:op_kernel_D2d}
&\sim \sigma_i \sigma_j \big[ (\tilde{\mathbf{l}}_i \cdot \tilde{\mathbf{m}}_j )  (\tilde{\mathbf{m}}_i \cdot \tilde{\mathbf{l}}_j ) 
+ (\tilde{\mathbf{l}}_i \cdot \tilde{\mathbf{l}}_j ) (\tilde{\mathbf{m}}_i \cdot \tilde{\mathbf{m}}_j )   \big]
(\tilde{\mathbf{n}}_i \cdot \tilde{\mathbf{n}}_j )  \nonumber \\
%& = [(l_a m_b + m_a l_b)n_c]_i [(l_a m_b + m_a l_b)n_c]_j \nonumber \\
& =  \mathrm{Tr }\bigg[ \big[ (\mathbf{l} \otimes \mathbf{m} + \mathbf{m} \otimes \mathbf{l}) \otimes \mathbf{n} \big]_i \cdot 
\big[ (\mathbf{l} \otimes \mathbf{m} + \mathbf{m} \otimes \mathbf{l}) \otimes \mathbf{n} \big]_j \bigg],
\end{align}
where the third-rank contraction $\mathrm{Tr}(\bullet_{abc} \cdot \bullet_{abc} )$ is determined up to the symmetries of the order parameter tensor (symmetric in the first two indices).
%where above in the second line, $(\sigma_i \sigma_j)^4 \equiv 1$ is inserted by which pseudo-vectorial degrees of freedoms are converted to vectorial ones.
By construction, the local quantity appearing in Eq.\eqref{eq:op_kernel_D2d} is $D_{2d}$ invariant, hence can be used to define a $D_{2d}$-biaxial order parameter,
\begin{equation} \label{eq:op_D2d_B}
\mathbb{B}^{D_{2d}} = (\mathbf{l} \otimes \mathbf{m} + \mathbf{m} \otimes \mathbf{l}) \otimes \mathbf{n}.
\end{equation}
%Moreover,  $\mathbb{B}^{D_{2d}}$ is traceless and we don't need take it trace out.
The full order parameter of a $D_{2d}$ nematic is therefore given by
\begin{equation} \label{eq:op_D2d}
\mathbb{O}^{D_{2d}}[\vek{l},\vek{m},\vek{n}] = \{ \mathbb{A}^{D_{\infty h}}[\vek{n}], \mathbb{B}^{D_{2d}}[\vek{l},\vek{m}, \vek{n}]\}.
\end{equation}

Continuing to $D_{4h}$ symmetry, after integrating out the gauge fields for $D_{4h}$, at the fourth order one will find the following contractions up to constant factors and terms solely depending on the axial axis $\mathbf{n}$,
\begin{align}
& \sim  \mathrm{Tr }\big[ (\mathbf{l}_i^{\otimes 4} + \mathbf{m}_i^{\otimes 4}) \cdot (\mathbf{l}_j^{\otimes 4} + \mathbf{m}_j^{\otimes 4}) 
+ 3 (\mathbf{l}_i^{\otimes 2}\otimes \mathbf{m}_i^{\otimes 2} 
\nonumber \\
& \quad + \mathbf{m}_i^{\otimes 2} \otimes \mathbf{l}_i^{\otimes 2}) \cdot (\mathbf{l}_j^{\otimes 2}\otimes \mathbf{m}_j^{\otimes 2} + \mathbf{m}_j^{\otimes 2}\otimes \mathbf{l}_j^{\otimes 2}) \big] .
\end{align}
One can therefore recognize two $D_{4h}$-invariant local tensors, 
\begin{align}
\mathbb{B}^{D_{4h}}_1 &= \mathbf{l}^{\otimes 2} \otimes \mathbf{m} ^{\otimes 2} + \mathbf{m}^{\otimes 2} \otimes \mathbf{l} ^{\otimes 2} 
   -\frac{4}{15} \delta_{ab} \delta_{cd} \bigotimes_{\substack{ \mu =  a,b,c, d}} \mathbf{e}_{\mu}
   \nonumber \\
 & \quad  + \frac{1}{15} \big(  \delta_{ac} \delta_{bd} \bigotimes_{\substack{ \mu =  a,c,b, d}} \mathbf{e}_{\mu}
    +  \delta_{ad} \delta_{bc}\bigotimes_{\substack{ \mu =  a,d,b, c}} \mathbf{e}_{\mu}  \big),    
    \\
 \mathbb{B}^{D_{4h}}_2 &= \mathbf{l}^{\otimes 4}  +  \mathbf{m}^{\otimes 4}  
-\frac{2}{15}\sum_{\text{pairs}} \delta_{ab} \delta_{cd} \bigotimes_{\substack{ \mu =  a,b,c, d}} \mathbf{e}_{\mu},
\end{align}
where we have subtracted the isotropic trace-part for convenience (``pairs" denotes the summation over all non-equivalent pairings of the indices of the Kronecker deltas).

However, these two tensors are not independent. Due the $O(3)$ relations Eq. \eqref{eq:O3_relation}, they satisfy
\begin{align}
\mathbb{B}^{D_{4h}}_1 + \mathbb{B}^{D_{4h}}_2 = (\vek{l}^{\otimes 2}+\vek{m}^{\otimes 2})^{\otimes 2} = (\id - \vek{n}^{\otimes 2})^{\otimes 2}.
\end{align}
This in turn means that both $\mathbb{B}^{D_{4h}}_1$ and $\mathbb{B}^{D_{4h}}_2$ have dependence on the axial axis $\mathbf{n}$.
Therefore, similar to the $D_{2h}$ case,  it is more convenient to use the linear combination $\mathbb{B}^{D_{4h}}_1 - \mathbb{B}^{D_{4h}}_2$ to characterize a $D_{4h}$ phase.

In case of $C_{6h}$ symmetry, the biaxial order parameters are rank-$6$ tensor and defined by the local contractions
\begin{align}
& \sim \mathrm{Tr} \big[  \mathbb{B}^{D_{6h}}_{1,i} \cdot \mathbb{B}^{D_{6h}}_{1,j} 
+ \mathbb{B}^{D_{6h}}_{2,i} \cdot \mathbb{B}^{D_{6h}}_{2,j}+
\mathbb{B}^{C_{6h}}_{1,i} \cdot \mathbb{B}^{C_{6h}}_{1,j} 
\nonumber \\
& \quad + \mathbb{B}^{C_{6h}}_{2,i} \cdot \mathbb{B}^{C_{6h}}_{2,j}
\big],
\end{align}
up to constant factors and terms depending on the axial axis $\mathbf{n}$, where the explicit form of these tensors are given in Table \ref{table:ops}.
The $D_{6h}$ order parameters appear here since $D_{6h}/ C_{6h} \simeq \{\id, c_2 (\mathbf{l})\}$ is a multiplicative group of order two acting trivially at even powers, leading to redundancy at even orders of the expansion Eq. \eqref{eq:k0_expansion}. The same phenomenon of course occurs for the $C_6$ quotients of $\{C_{6v}, D_{6}, D_{6h}\}$ etc. and the sixth order expansions coincide for the groups with identical order parameters.

Again due to the $O(3)$ relation Eq. \eqref{eq:O3_relation} and Eq.\eqref{eq:sigma}, these order parameters are not independent.
$\mathbb{B}^{D_{6h}}_{1}+\mathbb{B}^{D_{6h}}_2=(\vek{l}^{\otimes 2}+\vek{m}^{\otimes 2})^{\otimes 3} = (\id-\vek{n}^{\otimes 2})^{\otimes 3}$ depends solely on $\vek{n}$, and $\mathbb{B}^{C_{6h}}_{1} - \mathbb{B}^{C_{6h}}_2$ can be expressed as a function of the pseudovector $\vek{\widetilde{n}}$.
As a consequence, the linear combination  $\mathbb{B}^{D_{6h}}_1-\mathbb{B}^{D_{6h}}_2$ and $\mathbb{B}^{C_{6h}}_1+\mathbb{B}^{C_{6h}}_2$ are the appropriate in-plane order parameters for these symmetries.

The above procedure of deriving the biaxial order parameter is valid for all axial nematics with finite $n$-fold rotational symmetries. Naturally, the rank of the biaxial order parameter tensor increases with $n$ and becomes infinite when $n \rightarrow \infty$. 
This reflects the fact that a biaxial order parameter does not exist for phases with an in-plane $SO(2)$ symmetry, 
$\{ C_{\infty}, C_{\infty v}, C_{\infty h}, D_{\infty}, D_{\infty h}\}$.

\subsection{Polyhedral nematics}\label{sec:polyhedral}

Let us end by discussing the order parameters for the polyhedral groups.

The proper tetrahedral group $T$ can be generated by a two-fold rotation $c_2(\mathbf{n})$, as that in Eq. \eqref{eq:D2d_generator}, and a three-fold rotation acting as a cyclic permutation of $\{ \mathbf{l}, \mathbf{m}, \mathbf{n} \}$ given by
\begin{align}
c_3(\mathbf{l}+\mathbf{m}+\mathbf{n})= \left(
\begin{array}{ccc}
 0 & 1 & 0 \\
 0 & 0 & 1 \\
 1 & 0 & 0
\end{array}
\right).
\end{align}
These result in $12$ proper rotations that leave a tetrahedron embedded in a cube with normals $\vek{l},\vek{m}, \vek{n}$ invariant. After summing over gauge fields $U_{ij} \in T$ in the expansion Eq. \eqref{eq:k0_expansion}, one finds in the third order the following local contractions,
\begin{align}
 & \sim \sigma_i \sigma_j \mathrm{Tr} \big[ \sum_{\rm cyc} \tilde{\mathbf{l}}_i \otimes \tilde{\mathbf{m}}_i \otimes \tilde{\mathbf{n}}_i  
\cdot \sum_{\mathrm{cyc}}  \tilde{\mathbf{l}}_j \otimes \tilde{\mathbf{m}}_j \otimes \tilde{\mathbf{n}}_j \big]
%& = \frac{1}{3} \mathrm{Tr} \big[\big( \sum_{\mathrm{cyc}}  \mathbf{l} \otimes \mathbf{m} \otimes \mathbf{n} \big)_i 
%\cdot \big( \sum_{\mathrm{cyc}}  \mathbf{l} \otimes \mathbf{m} \otimes \mathbf{n} \big)_j \big],
\end{align}
where $\sum_{\mathrm{cyc}}$ runs over cyclic permutations of $\{ \mathbf{l}, \mathbf{m}, \mathbf{n} \}$.
Hence we can define the $T$-invariant local tensor,
\begin{align} \label{eq:T_op}
\mathbb{O}^T  = \mathbb{O}^T_1=  \sum_{\mathrm{cyc}}  \mathbf{l} \otimes \mathbf{m} \otimes \mathbf{n}.
\end{align}

$\mathbb{O}^T$ in Eq. \eqref{eq:T_op} contains only cyclic permutations of the three local axes and carries a chirality, as there are no improper operations in $T$.
By interchanging two of these axes corresponding to a reflection, we obtain an equivalent $T$-invariant tensor but with different handedness,
\begin{align} \label{eq:T_op_2}
\mathbb{O}^{T}_2 =  \sum_{\mathrm{cyc}}  \mathbf{m} \otimes \mathbf{l} \otimes \mathbf{n}.
\end{align}
One realizes that a linear combination of $\mathbb{O}^T_1$ and $\mathbb{O}^{T}_2 $ will give an ordering tensor that is invariant under the symmetry group of a regular tetrahedron, $T_d$.
Indeed, integrating out the gauge fields for the $T_d$ group, where $T_d = T \rtimes \{\id , \sigma_d \}$ and $\sigma_d$ defined in Eq. \eqref{eq:D2d_generator}
generates the odd permutation, one will find in the third order of Eq. \eqref{eq:k0_expansion}
\begin{align}
 \sim \mathrm{Tr} \bigg[ \sum_{\rm cyc}\big(\mathbf{l}_i \otimes \mathbf{m}_i + \mathbf{m}_i \otimes \mathbf{l}_i )\otimes \mathbf{n}_i  \nonumber\\
\cdot \sum_{\mathrm{cyc}} 
 (\mathbf{l}_j \otimes \mathbf{m}_j + \mathbf{m}_j \otimes \mathbf{l}_j )\otimes \mathbf{n}_j \bigg]
\end{align}
giving precisely the order parameter tensor 
\begin{align} \label{eq:Td_op}
\mathbb{O}^{T_d} =  \sum_{\mathrm{cyc}}  ( \mathbf{l} \otimes \mathbf{m} +  \mathbf{m} \otimes \mathbf{l})\otimes \mathbf{n}
\end{align}
as expected (compare to the $D_{2d}$ third-rank order parameter).

There is yet another point group belonging to the tetrahedral group family, the group $T_h = T\times \{\id, -\id\}$.
Interestingly, due to $T_h = T \times \{ \id, -\id \}$, all odd orders in the expansion Eq. \eqref{eq:k0_expansion} vanish and the first non-trivial terms appear in the fourth order with the contractions,
\begin{align} \label{eq:expansion_Th}
 \sim &\big( \mathbf{l}_i^{\otimes 4}  +  \mathbf{m}_i^{\otimes 4} + \mathbf{n}_i^{\otimes 4} \big)  \cdot  \big( \mathbf{l}_j^{\otimes 4}  +  \mathbf{m}_j^{\otimes 4} + \mathbf{n}_j^{\otimes 4} \big)  \nonumber \\
&+ \big( \mathbf{l}_i^{\otimes 2} \otimes \mathbf{m}_i^{\otimes 2} 
+  \mathbf{m}_i^{\otimes 2} \otimes \mathbf{n}_i^{\otimes 2} 
 +  \mathbf{n}_i^{\otimes 2} \otimes \mathbf{l}_i^{\otimes 2}\big)
 \nonumber \\
&\phantom{+} \cdot  \big( \mathbf{l}_j^{\otimes 2} \otimes \mathbf{m}_j^{\otimes 2} \! +\!  \mathbf{m}_j^{\otimes 2} \otimes \mathbf{n}_j^{\otimes 2}  +   \mathbf{n}_j^{\otimes 2} \otimes \mathbf{l}_j^{\otimes 2}\big). 
\end{align}
The second term in the above expression gives the $T_h$ invariant order parameter tensor
\begin{align} \label{eq:op_Th}
\mathbb{O}_1^{T_h}
&=  \mathbf{l}^{\otimes 2} \otimes \mathbf{m}^{\otimes 2} +  \mathbf{m}^{\otimes 2} \otimes \mathbf{n}^{\otimes 2} +  \mathbf{n}^{\otimes 2} \otimes \mathbf{l}^{\otimes 2}
\nonumber \\
 & \quad -\frac{2}{5} \delta_{ab} \delta_{cd}\bigotimes_{\substack{ \mu =  a,b,c, d}} \mathbf{e}_{\mu}
+ \frac{1}{10} \big( \delta_{ac} \delta_{bd} \bigotimes_{\substack{ \mu =  a,c,b, d}} \mathbf{e}_{\mu}
\nonumber \\   
& \quad    + \delta_{ad} \delta_{bc}\bigotimes_{\substack{ \mu =  a,d,b, c}} \mathbf{e}_{\mu} \big)  ,       
\end{align}
where we have subtracted the trace. The first term in Eq. \eqref{eq:expansion_Th} actually coincides with the $O_h$ ordering tensor $\mathbb{O}^{O_h}$.
Again, this is because $O_h/T_h \simeq \{\id, \sigma_d\}$ is a group of order two acting trivially on $\mathbb{O}^{O_h}$ that leads to some redundancy at even orders in the expansion. Moreover,
$\mathbb{O}_1^{T_h}$ in Eq. \eqref{eq:op_Th} is not invariant under interchanging $\mathbf{l}$ and $\mathbf{m}$, which corresponds to the four fold rotation in $O_h$.
Therefore, we can define another $T_h$-invariant tensor, 
\begin{align}
\mathbb{O}^{T_h}_2 &=    \mathbf{m}^{\otimes 2} \otimes \mathbf{l}^{\otimes 2} +  \mathbf{n}^{\otimes 2} \otimes \mathbf{m}^{\otimes 2} +  \mathbf{l}^{\otimes 2} \otimes \mathbf{n}^{\otimes 2}
\nonumber \\
 \nonumber \
 & \quad -\frac{2}{5} \delta_{ab} \delta_{cd}\bigotimes_{\substack{ \mu =  a,b,c, d}} \mathbf{e}_{\mu}
+ \frac{1}{10} \big( \delta_{ac} \delta_{bd} \bigotimes_{\substack{ \mu =  a,c,b, d}} \mathbf{e}_{\mu}
\nonumber \\   
& \quad    + \delta_{ad} \delta_{bc}\bigotimes_{\substack{ \mu =  a,d,b, c}} \mathbf{e}_{\mu} \big)  .  
\end{align}
Due to the $O(3)$ constraints, however, this and the two terms in Eq. \eqref{eq:expansion_Th} are not independent,
\begin{align}
& \mathbb{O}^{O_h} + \mathbb{O}^{T_h}_1 + \mathbb{O}^{T_h}_2 \nonumber \\
&= (\vek{l}\otimes\vek{l} + \vek{m}\otimes\vek{m}+\vek{n}\otimes\vek{n})^{\otimes 2} + \rm{const.} \nonumber \\
&= \id \otimes \id + \rm{const.}
\end{align}
Therefore, both $\mathbb{O}^{T_h}_1 $ and $\mathbb{O}^{T_h}_2$ suffice to describe the $T_h$ orientational order.

Proceeding to $O$ and $O_h$ group, non-zero terms appear likewise in fourth order of the expansion and now one will obviously find the same contraction as the first term in Eq. \eqref{eq:expansion_Th} up to a constant factor, hence one can define the $O_h$ order parameter tensor as
\begin{align}
\mathbb{O}^{O_h} =  \mathbf{l}^{\otimes 4}  +  \mathbf{m}^{\otimes 4} + \mathbf{n}^{\otimes 4} 
-\frac{1}{5}\sum_{\text{pairs}} \delta_{ab} \delta_{cd} \bigotimes_{\substack{ \mu =  a,b,c, d}} \mathbf{e}_{\mu}. 
\end{align}
For the proper subgroup $O$, we have an additional non-trivial third order in the expansion, which is simply $\sim \sigma_i \sigma_j$ giving the chiral order parameter.

The same procedure discussed above applies as well for $I$ and $I_h$ group and leads to a rank-$6$ ordering tensor,
\begin{align} \label{eq:op_Ih}
\mathbb{O}^{I_h} &= \sum_{\rm{cyc}} \big[ \mathbf{l}^{\otimes 6} +\sum_{ \{+,- \}} 
			\big(\frac{1}{2} \mathbf{l} \pm \frac{\tau}{2} \mathbf{m} \pm \frac{1}{2\tau} \mathbf{n} \big)^{\otimes 6} \big]
\nonumber \\		
  		& \quad -\frac{1}{7} \sum_{\text{pairs}} \delta_{ab} \delta_{cd} \delta_{ef} \bigotimes_{\substack{ \mu =  a,b,c, \\  d,e,f}} \mathbf{e}_{\mu} .
\end{align}
Here we express $\mathbb{O}^{I_h}$ in a compact form as the sixth-order tensors product of 15 vectors that was also discussed in Ref. \cite{Fel1995}, bearing in mind that due to the high rank and the five-fold rotations in $I_h$, it contains numerous terms that are practically very unwieldy. 

To stress the advantage of our gauge theory formulation \cite{LiuEtAl2015b} even more explicitly, we note that the relation between, say, Eq. \eqref{eq:op_Ih} and its fully expanded form is not in principle explicitly needed in the gauge theoretical formulation but the order parameter arises by construction from the much more manageable Eq. \eqref{eq:gauge theory}.%, as we have demonstrated.

\section{Summary and outlook}\label{sec:summary}

The rotational symmetries of three dimensional isotropic space $O(3)$ can in principle break to any non-trivial point group. According to the Landau-de Gennes paradigm, each symmetry is accompanied by a order parameter and associated phase transitions. These order parameters are high-rank tensors and quite involved in general. 
As result, this remaining frontier of phenomenological Landau-de Gennes order parameter theory has remained rather elusive, apart from the extensively discussed and experimentally realized cases of uniaxial and biaxial nematics. However, also the challenge of the experimental realization of generalized nematics might be overcome in the near future in view of the rapid experimental advances in the availability and control of new promising platforms for generalized nematic phases \cite{DamascenoGlotzer2012, vanAndersGlotzer2014, Glotzer2007, SunXia2002, HuangMichael2012, Mark2013}. 

In this paper we have completed the first step towards bridging this theoretical gap by presenting the order parameter tensors related to the orientational ordering with unbroken three-dimensional point-group symmetries. 
Perhaps surprisingly, this is to the best of our knowledge the first time these have appeared in such a generality. In the context of a gauge theoretical lattice model, we have developed a systematic way of classifying the minimal order parameter tensos of arbitrary 3D point group symmetries and have presented the explicit form of these order parameters for an extensive selection of the physically most relevant symmetries. Although we arrived to these results utilizing the gauge theoretical lattice model we constructed earlier for the study of generalized nematics \cite{LiuEtAl2015b}, the results are of course independent of the gauge theoretical machinery. In addition to our lattice model, with the order parameters it is in principle possible to study the nematic phases via Landau-de Gennes theories by considering all symmetry allowed couplings of the order parameters, for example using the approach outlined in Ref. \onlinecite{LubenskyRadzihovsky2002}.
Given the universality of the applications of the orientational tensor order parameters our work is of general interest across many different fields, in particular we anticipate that our results can provide for a road map for the search of new nematic phases of matter. \\

%\section{Acknowledgments}

\textbf{Acknowledgments} 
We would like to thank E. Cobanera, G. T. Barkema and D. Kraft for helpful dicussions. This work has been supported by the Dutch Foundation on the Research of Fundamental Matter (FOM), which is part of NWO. K. L. is supported by the State Scholarship Fund program organized by China Scholarship Council (CSC). K. W. is supported by DOE-BES Division of Materials Sciences and Engineering DMSE at Stanford University.

\bibliographystyle{apsrev4-1}
\bibliography{nematics}

\end{document}